\documentclass[aps,pra,a4paper,showpacs,twocolumn,10pt]{revtex4-1}
\usepackage[utf8x]{inputenc}    %
\usepackage{amssymb,amsmath}    %
\usepackage{graphicx}           %
\usepackage{textcomp}           %
\usepackage{color}              %
\usepackage{epstopdf}           %
\usepackage{bbm}                %
\usepackage{bm}			 %
\usepackage{mathtools}		 %
\usepackage{xfrac}
\usepackage[caption=false]{subfig}		%
\usepackage{geometry}
\definecolor{myurlcolor}{rgb}{0,0,0.7}
\definecolor{myrefcolor}{rgb}{0.8,0,0}
\usepackage[unicode=true,pdfusetitle, bookmarks=false,bookmarksnumbered=false,
bookmarksopen=false, breaklinks=false,pdfborder={0 0 0},backref=false,
colorlinks=true, linkcolor=myrefcolor,citecolor=myurlcolor,urlcolor=myurlcolor]
{hyperref}

\newcommand{\bra}[1]{\mathinner{\langle #1\rvert}}
\newcommand{\ket}[1]{\mathinner{\lvert#1\rangle}}
\newcommand{\Bra}[1]{\left< #1 \right|}
\newcommand{\Ket}[1]{\left| #1 \right>}
\newcommand{\ketbra}[2]{\mathinner{\lvert#1\rangle\langle #2\rvert}}
\newcommand{\Ketbra}[2]{\left|#1\middle>\middle<#2\right|}

\newcommand{\sketbra}[3]{\ket{#1}_{#3}\!\bra{#2}}
\newcommand{\Sketbra}[3]{\Ket{#1}_{#3}\!\Bra{#2}}

\definecolor{orange}{RGB}{255,165,0}

\begin{document}
\title{Measurement based quantum communication with resource states generated by entanglement purification}
 \date{\today}
 \author{J. Wallnöfer}
 \author{W. Dür}
 \affiliation{Institut für Theoretische Physik, Universität Innsbruck, Technikerstr. 21a, A-6020 Innsbruck, Austria}
 \begin{abstract}
We investigate measurement-based quantum communication with noisy resource states that are generated by entanglement purification. We consider the transmission of encoded information via noisy quantum channels using a measurement-based implementation of encoding, error correction and decoding. We show that such an approach offers advantages over direct transmission, gate-based error correction and measurement-based schemes with direct generation of resource states. We analyze the noise structure of resource states generated by entanglement purification and show that a local error model, i.e. noise acting independently on all qubits of the resource state, is a good approximation in general, and provides an exact description for $GHZ$-states. The latter are resources for a measurement-based implementation of error correction codes for bit-flip or phase flip errors. This provides a first link between the recently found very high thresholds for fault-tolerant measurement-based quantum information processing based on local error models for resource states, to error thresholds for gate based computational models.

 \end{abstract}

 \maketitle

 \section{Introduction \label{sec:intro}}
 Quantum communication is a key aspect to make the tools of quantum information processing available for spatially
  separated parties. Quantum error correction \cite{niechu,clusterringcode} and entanglement purification \cite{bbpssw, dejmps, eppreview} are two
  approaches to combat the noise and imperfections one invariably encounters in any realistic setting, and it are these
  methods that make sharing entanglement over large distances feasible.

  The measurement based approach to quantum computation \cite{natphys_mbqc, oneway, tqc} is one way to implement the protocols used in quantum communication
  to establish long distance entanglement.  Recent results \cite{mbqrepeaters,mbqc_eppthresh,hybrid,zwerger_review} show that a measurement based approach can be applied also in the context of entanglement purification, quantum error corrections, quantum communication and hybrid quantum computation, and offers high error thresholds. It was found that a measurement based implementation can tolerate more than
  23\% local depolarizing noise per qubit for bipartite entanglement purification \cite{mbqc_eppthresh} and more than
  13\% for fault-tolerant quantum computation using a hybrid model \cite{hybrid}. While these thresholds are the limits of
  these approaches, they highlight the application potential of small scale resource states and provide motivation to study
  them further. Elements of such a measurement-based approach to quantum error correction have already been experimentally demonstrated with trapped ions \cite{exp_lanyon} and photons \cite{exp_barz}. Furthermore, for entanglement purification at the logical level \cite{logic_purify} a measurement based implementation would be especially suitable. 

  Here we investigate multipartite entanglement purification \cite{prlmepp,adbepp,eppallgraphs,eppreview,GCRmepp,GKVmepp} as one specific approach to generate
  resource states for quantum error correction and quantum communication with a high fidelity. We analyze the performance of states generated in this way in quantum communication protocols that work via the transmission of encoded quantum information using an error correction code. Note that entanglement purification can both be used to generate entangled states between distant parties as well as at individual sites to generate resource states to locally implement a quantum task in a measurement based way. We also investigate the noise structure of the purified states, as the locality of the noise is one important assumption to obtain the thresholds mentioned above \cite{zwerger_review}. While it has been conjectured that the states resulting from imperfect entanglement purification should approximately be described by local noise \cite{lnconjecture}, this has not been formally proven or investigated.

  Our two main results are as follows:
  \begin{itemize}
  \item A local noise model is a good approximation to describe resource states generated by imperfect entanglement purification. In general, this description is not exact, however the amount of non-local noise is very small.  For certain resource states such as Greenberger-Horne-Zeilinger (GHZ) states, an exact description by a local noise model is possible by slightly reducing the fidelity.
  \item Encoded quantum communication over noisy channels where encoding, decoding and error correction is implemented in a measurement based way with resource states generated by entanglement purification outperforms direct transmission of quantum information, schemes making use of gate-based implementation of error correction as well as measurement-based schemes where resource states are generated directly via gates.
  \end{itemize}
  Furthermore, for a specific, restricted noise process and task, we provide a connection between error models for measurement-based implementation of quantum error correction
  (where one considers noisy resource states) on the one hand, and gate-based error models on the other hand. This link is of particular interest as the error thresholds for measurement-based approaches
  have been shown to be an order of magnitude higher than for gate-based approaches \cite{mbqrepeaters,mbqc_eppthresh,hybrid,zwerger_review}, but the underlying error models
  are not directly comparable.

  The paper is organized as follows: In section \ref{sec:concepts} we give an overview of the concepts and notations used. Reader familiar with graph states, error correction, measurement-based quantum information processing or multipartite entanglement purification may skip the corresponding sections.
  In section \ref{sec:noisestruct} we investigate whether the noise of the state generated by an entanglement purification protocol (EPP)
  can be described by local noise for GHZ states and the resource state for the cluster-ring code. We also show that noise can always be localized for GHZ states.
  In section \ref{sec:application} we analyze the performance of the resource states generated by entanglement purification in a quantum communication setting.  In section \ref{sec:scaling} we provide a connection between measurement-based and gate-based error models for a simplified error model. To this aim we investigate the scaling behavior with the number of encoding qubits for the measurement based repetition code.

 \section{Background \label{sec:concepts}}
  Here we review the concepts used in this work as well as introducing some notation. We explain the errors model used
  for our analysis and review graph states, a special class of multipartite quantum states that we use in the context
  of measurement based quantum error correction. Multipartite entanglement purification is one way to generate
  high fidelity graph states that is also discussed here.

 \subsection{Error model \label{sec:errormodel}}
Noise channels are modeled by completely positive maps. We consider
local Pauli diagonal noise channels acting on the $i$-th qubit, described by
 \begin{equation}
  \mathcal{E}^i(\vec{p}) \rho = p^0 \mathbbm{1} \rho \mathbbm{1} + p^1 \sigma_x^i \rho \sigma_x^i +
  p^2 \sigma_y^i \rho \sigma_y^i + p^3 \sigma_z^i \rho \sigma_z^i
 \end{equation}
  with the error parameters $\vec{p}$ that have to satisfy: $p^0 + p^1 + p^2 + p^3 = 1$.
 Some special channels of this kind include the local bit-flip channel ($\sigma_x$ noise) $\mathcal{D}_x (p_x)$
 with $\vec{p}=[p_x,(1-p_x),0,0]$,
 the local phase-flip channel ($\sigma_z$ noise) $\mathcal{D}_z (p_z)$ with $\vec{p}=[p_z,0,0,(1-p_z)]$
 and the local depolarizing (white noise channel $\mathcal{D}_w (p)$ with $\vec{p}=[p+(1-p)/4,(1-p)/4,(1-p)/4,(1-p)/4]$.

 For the gate based protocols we assume that local Clifford operations can be done perfectly, but that the two-qubit entangling gates like the CNOT gate $U^{1\rightarrow 2}_\mathrm{CNOT} = \Ketbra{0}{0} \otimes \mathbbm{1} + \Ketbra{1}{1} \otimes \sigma_x$ are noisy. The CNOT gate is then described by the map $\mathcal{M}_\mathrm{CNOT}$ which is modeled by local noise channels $\mathcal{D}$ acting on the source and the target qubit followed by the perfect operation
  \begin{equation}
 \mathcal{M}^{1 \rightarrow 2}_{\mathrm{CNOT}}(p) \rho = U^{1 \rightarrow 2}_{\mathrm{CNOT}} \left( \mathcal{D}^{1}(p)
  \mathcal{D}^{2}(p) \rho \right) U^{\dagger \ 1 \rightarrow 2}_{\mathrm{CNOT}} \text{.}
  \label{eqn:mcnot}
 \end{equation}
For the measurement based implementations we assume that the imperfections arise from only having access to imperfect resource states
and external sources but the Bell measurements for the read-in can be done perfectly. One may also consider imperfect Bell measurements that can be modelled by local depolarizing noise, followed by a perfect measurement. In this case, noise of Bell measurements can be included in the external error parameter.

Occasionally we will also refer to the 2-qubit depolarizing noise
 \begin{equation}
 \mathcal{D}^{a,b}_w (p') \rho = p' \rho + \frac{1-p'}{4} \mathrm{tr}_{a,b}(\rho) \otimes \mathbbm{1}^{a,b}
 \end{equation}
and the global depolarizing noise channel
 \begin{equation}
  \mathcal{D}_g (\widetilde{p}) \rho = \widetilde{p} \rho + \frac{1-\widetilde{p}}{2^N} \mathbbm{1} \text{.}
 \end{equation}

 \subsection{Graph states \label{sec:graph}}
 A simple, unweighted and undirected graph $G$ with $N$ vertices is described by the pair $G=(V,E)$.
 $V = \left\{ 1, 2, \dots , N \right\}$ is the set of vertices and $E \subseteq [V]^2$
 is the set of edges, where $[V]^2$ is the set of subsets of $V$ containing $2$ elements.

 Graph states \cite{firstgraphstate, ent_graphstates, graphstates} are a special class of states which can be represented by a graph in such a way
 that each vertex corresponds to a qubit and every edge indicates a two-body interaction
 between the qubits it connects.
 The graph state $\Ket{G} \in (\mathbbm{C}^2)^{\otimes N}$ corresponding to the graph $G$ is given by
 \begin{equation}
  \Ket{G} = \prod_{ \left\{a,b \right\} \in E } U_\mathrm{CZ}^{ab} \Ket{+}^{\otimes N}
  \label{eqn:graphstate}
 \end{equation}
 where $U_\mathrm{CZ}^{ab}=\Ketbra{0}{0} \otimes \mathbbm{1} + \Ketbra{1}{1}\otimes \sigma_z$
 is the controlled-$\sigma_z$ gate acting on qubits $a$ and $b$, and $\Ket{+}$ is the eigenstate of $\sigma_x$ with eigenvalue $+1$.

 The orthogonal graph state basis is defined as
 \begin{equation}
  \Ket{\bm{\mu}}_G = \prod_{j \in V} \left( \sigma_z^j \right)^{\mu_j} \Ket{G}
 \end{equation}
 with $\bm{\mu}=(\mu_1, \mu_2, \dots, \mu_N) \in \{0,1\}^N$. The subscript $G$
 acts as a reminder that these states are defined with respect to a particular graph
 and to distinguish them from product states in the computational basis.

 These states can be uniquely identified by the eigenvalues of the $N$ correlation
 operators $K_a = \sigma_x^a \prod_{\{a,b\} \in E} \sigma_z^b$ for $a \in V$
 \begin{equation}
  K_a \Ket{\bm{\mu}}_G = (-1)^{\mu_a} \Ket{\bm{\mu}}_G \text{.}
 \end{equation}
 A graph is called $k$-colorable if the vertices of the graph can be colored in using
 $k$ colors in such a way that no two vertices with the same color are connected by an edge.

\begin{figure}[ht]
  \centering
  \subfloat[\centering \label{fig:ghz4}]{\includegraphics[width=0.33\columnwidth]{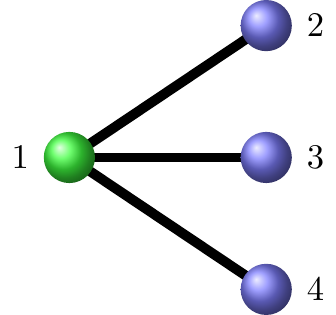}  }
  \hfill
  \subfloat[\centering \label{fig:clusterringcode}]{\includegraphics[width=0.48\columnwidth]{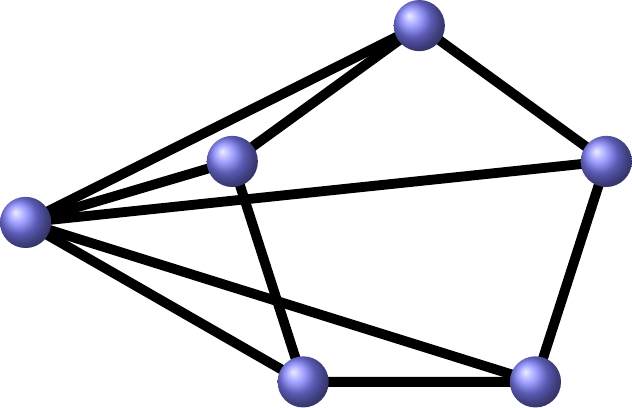}  }
   \caption{(a) Graph state version of the GHZ state with $N=4$. The qubit in set A is colored green, qubits in set B are colored blue.
	    (b) Resource state for measurement based implementation of the cluster-ring quantum error correction code. \label{fig:graphexamples} }
  \end{figure}

 The graphs in figure \ref{fig:graphexamples} are examples of useful graph states.

 \subsection{Quantum error correction \label{sec:qec}}
 Quantum error correction is one approach to deal with the imperfections and noise effects that
 invariably occur in any realistic setting for quantum information processing tasks.
 It relies on encoding information in a higher dimensional Hilbert space in a way that errors map
 the logical state on orthogonal subspaces, such that errors can be detected and corrected. We will consider the following two
 error correction codes:

 The repetition code is a simple way to encode information. The simplest example is given by the three qubit bit-flip code (see e.g. \cite{niechu}). The state of a qubit $\Ket{\Psi} = \alpha \Ket{0} + \beta \Ket{1}$
 is encoded as $\alpha \Ket{000} + \beta \Ket{111}$ using three qubits. This allows one to detect
 if a $\sigma_x$ error occurs on one of the qubits and apply an appropriate correction operation. The code is easily generalized to protect against $m$ bit-flip errors by using $2m+1$ qubits.

 The five qubit cluster ring code \cite{clusterringcode} is an example of an optimal code to correct
 arbitrary single-qubit errors. Its logical zero state $\Ket{0_L}$ is given by the graph state corresponding
 to a closed linear cluster with five qubits (as depicted in figure \ref{fig:clusterringcode} without the read-in qubit on the left)
 \begin{equation}
   \Ket{0_L} = \Ket{G} = \Ket{00000}_G \; ; \;
   \Ket{1_L} = \Ket{11111}_G \text{.}
 \end{equation}
 Each type of single-qubit noise produces a distinct pattern, which means that each error operator map the initial logical subspace to a distinct, orthogonal subspace.
 This can be seen by using the rules for Pauli noise acting on graph states \cite{ent_graphstates,graphstates}.

 When discussing quantum error correction codes, the parameter ranges of the noise channels for which it is beneficial to use error correction
 are of central importance. The error channels acting on the physical qubits that are used to encode a state translate to some error probability
 on the logical level after correction \cite{effectivenoise}. As long as the error rate on the logical level is smaller than on the physical level, using the error correction code is beneficial as compared to no encoding. For the three qubit bit-flip code with $\mathcal{D}_x(p_x)$ acting on each of the encoded qubits
 the threshold is $p_x = 0.5$, while the 5-qubit code described above tolerates local white noise $\mathcal{D}_w(p)$ up to p=0.8250 \cite{5qubitthresh}.

 While being above these thresholds ensures that the information decoheres slower, by using these codes in a concatenated way it is possible to achieve
 unit fidelity as the error can be reduced exponentially with only polynomial overhead \cite{5qubitthresh, effectivenoise}.

 \subsection{Measurement based quantum computation \label{sec:mbqc}}
 In contrast to other models of quantum computation, the central idea of measurement based quantum computation \cite{natphys_mbqc, oneway} is to use entangled states as a resource  to process quantum information. This can be accomplished by measurements on the resource states only.
 The particular approach used in this work can either be understood as application of one-way quantum computation to particular task \cite{firstgraphstate},
 or as teleportation-based computation \cite{teleportation,tqc}.

 The key notion is to use a modified teleportation protocol which instead of Bell pairs uses a different resource state in order to implement
 a quantum operation on the input pairs. The resource state implementing a desired map $\mathcal{M}^{12\dots n}$ acting on $n$ qubits can be found using the Choi-Jamiolkowski
 isomorphism \cite{jamiolkowski}. The mixed state $\rho_\mathcal{M}$ that can be used to implement this map probabilistically is given by $n$ copies of the pure state $\Ket{\Phi^+}$
 with the map acting on one of the halves for each of the copies
  \begin{equation}
  \label{rho_jam}
   \rho_\mathcal{M}=\mathcal{M}^{B_1 B_2 \dots B_n} \bigotimes_{i=1}^n \Sketbra{\Phi^+}{\Phi^+}{A_iB_i} \text{.}
  \end{equation}
 Additional qubits in the Jamiolkowski state that correspond to fixed inputs or auxiliary qubits being measured can be treated as virtual particles
 that are useful to understand the construction of the resource state but do not need to be prepared physically.
 
 Similar as in the teleportation protocol, Pauli correction operations depending on the outcome of Bell measurements are applied to the input state prior to the application of the map.
 For example consider a one qubit unitary operation $U$ that is supposed to be applied on a input state $\Ket{\psi}_{A'}$ using the resource state $\mathbbm{1} \otimes U \Ket{\Phi^+}_{AB}$.
 If the outcome of the Bell measurement on $A$ and $A'$ is $\Ket{\Psi^+} = \mathbbm{1} \otimes \sigma_x \Ket{\Phi^+}$ one obtains
 \begin{equation}
  \sketbra{\Psi^+}{\Psi^+}{AA'} \ket{\psi}_{A'} \left(\mathbbm{1} \otimes U \ket{\Phi^+}_{AB}\right) \propto \ket{\Psi^+}_{AA'} (U \sigma_x \ket{\psi}_B )
 \end{equation}
 so $U \sigma_x$ is implemented instead of the desired unitary $U$.
 
 In general these Pauli by-product operators cannot be undone, yielding a probabilistic implementation of the desired map. A Clifford circuit can however be implemented deterministically with a resource state of minimal size, that is with only input and output qubits, because
 the byproduct Pauli operators that appear from the random outcomes of the Bell measurements can be permuted with the map and it is always
 possible to correct them \cite{firstgraphstate, hybrid, mbqrepeaters}. So in the example above, if $U$ is a Clifford operation we know that $U\sigma_x = \sigma_k U$ with $k \in \{x,y,z\}$,
 so we can simply apply $\sigma_k$ to the output state and obtain the desired result $U\ket{\psi}$.
 
 Additional qubits in the Jamiolkowski state that correspond to fixed inputs or auxiliary qubits being measured can be treated as virtual particles that are useful to understand the construction of the resource state but do not need to be prepared physically.

 For non-Clifford operations the Pauli operators cannot be permuted with the map. Hence one either has to be content with a probabilistic implementation of the operation, or one has to give up the requirement of minimal size as additional auxiliary particles allow one to make the process deterministic by using sequential measurements \cite{oneway}. That means one uses resource states that have additional qubits, which are measured only after the read-in with Bell measurements has been performed since the measurement directions have to be adjusted depending on the outcomes.

 Both, entanglement purification and quantum error correction, the protocols considered here, use only Clifford operations and can therefore
 be implemented deterministically with resource states of minimal size \cite{zwerger_review}. 
 
\begin{figure}
  \includegraphics[width=0.8\columnwidth]{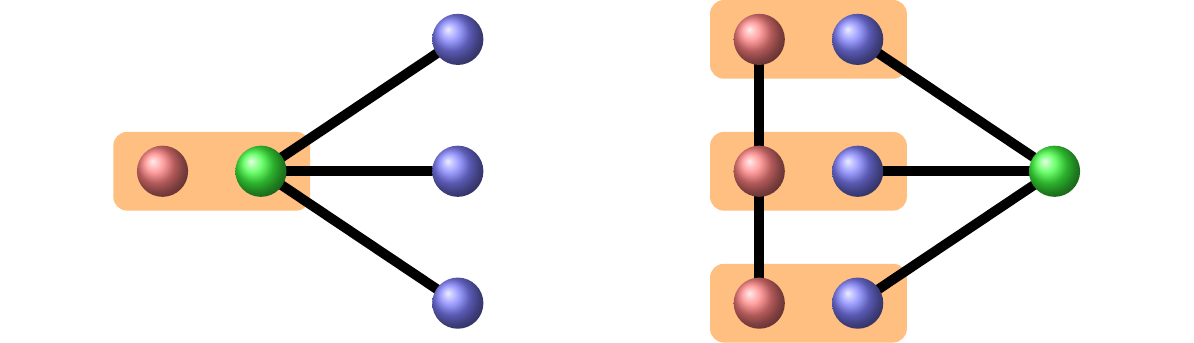}
  \caption{\label{fig:encdecexample} Measurement based implementation of encoding (left) and decoding (right) for the 3-qubit repetition code using a GHZ state. The boxes indicate
	    Bell measurements.}
 \end{figure}

 The overall setting of a measurement-based implementation of different quantum information processing tasks is shown in figure \ref{fig:encdecexample}.
 The quantum information in one qubit can be encoded using the appropriate resource state and performing the read-in using a Bell measurement. Depending on the
 outcome one also has to apply local Pauli operations as correction similar to the correction for the by-product operators in a teleportation protocol.
 The decoding can also be done using the same state, but switching the roles of input and output qubits, as depicted in the right part of figure \ref{fig:encdecexample}.
 The outcomes of the Bell measurements in this case do not only include information about the by-product operators inherent to the measurement based approach
 but also the error syndrom revealed by the error correction code and can therefore be used to derive the appropriate correction operator (see Appendix \ref{sec:dec_patterns} for details).
 
 \subsection{Multipartite entanglement purification \label{sec:epp}}
 Entanglement purification \cite{bbpssw,dejmps,adbepp,eppallgraphs,eppreview,prlmepp,GCRmepp,GKVmepp}
 is a method to counteract the negative effects of noise
 on entangled states. The key idea is to use several copies of a noisy entangled
 state and apply a protocol that produces fewer, but less noisy copies, using only
 local operations and classical communication (LOCC).

 We briefly discuss a (multipartite) entanglement purification protocol (EPP) that works on two-colorable
 graph states \cite{prlmepp,adbepp}. We start from a density operator diagonal in the
 graph state basis corresponding to the desired graph state
\begin{equation}
 \rho = \sum_{\bm{\mu}_A,\bm{\mu}_B} \lambda_{\bm{\mu}_A,\bm{\mu}_B} \Sketbra{\bm{\mu}_A,\bm{\mu}_B}{\bm{\mu}_A,\bm{\mu}_B}{G}
\end{equation}
 with the binary vector index $\bm{\mu}$ split in two parts $\bm{\mu}_A$ and $\bm{\mu}_B$ to emphasize the
 two sets of qubits corresponding to different colors. It suffices to consider states diagonal in this basis as every state can be brought
 to this form via depolarization \cite{adbepp}.
 This protocol consists of two subprotocols P1 and P2.
 Protocol P1 starts with applying CNOT operations on two copies $\rho_\mu$ and $\rho_\nu$ of the same state.
 For qubits in set A the second copy $\rho_\nu$ is used as the source and the first copy $\rho_\mu$ as the target of the CNOTs
 while for qubits in set B the CNOTs are applied the other way around.
 This multilateral CNOT operation has the effect
 \begin{equation}
  \Ket{\bm{\mu}_A,\bm{\mu}_B}_G \Ket{\bm{\nu}_A,\bm{\nu}_B}_G \rightarrow
  \Ket{\bm{\mu}_A, \bm{\mu}_B \oplus \bm{\nu}_B}_G \Ket{\bm{\nu}_A \oplus \bm{\mu}_A, \bm{\nu}_B}_G \text{.}
 \end{equation}
 Then all correlation operators $K_{\bm{\nu}_A}$ belonging to set A are measured on the second copy. This can be accomplished by local, single-qubit measurements. The protocol is successful only if all measurement outcomes are $+1$, otherwise it failed and the states have to be discarded.
 If this purification step is succesful the coefficients of the density operator of the first copy have been
 updated
  \begin{equation}
   \lambda'_{\bm{\gamma}_A,\bm{\gamma}_B} = \quad \sum_{\mathclap{\{ (\bm{\mu}_B, \bm{\nu}_B)|\bm{\mu}_B\oplus \bm{\nu}_B = \bm{\gamma}_B\}}} \quad \frac{\lambda_{\bm{\gamma}_A,\bm{\mu}_B} \lambda_{\bm{\gamma}_A,\bm{\nu}_B}}{2K} \text{,}
  \end{equation}
  where the denominator is a normalization constant and the probability of success.
 Generally the effect of P1 is to amplify coefficients with $\bm{\mu}_A = \bm{0}$.

 Subprotocol P2 works just like protocol P1 but the role of sets A and B are switched. Applying both
 protocols in an alternating way allows us to amplify only the desired coefficient $\lambda_{\bm{0},\bm{0}}$.

 There also exist EPPs that works on all graph states \cite{eppallgraphs} that is based on similar principles but needs additional
 auxiliary states corresponding to different two-colorable graphs. This protocol is used to purify the resource state for the
 cluster-ring code (Appendix \ref{sec:purdetails}).

 \section{Noise structure of states generated by entanglement purification \label{sec:noisestruct}}
 We are interested in the noise structure of the state at the fixed point of an EPP,
 especially whether that state can be described by a local noise model. This is of particular interest because
 local noise on the resource state has been used as a model for analysing measurement based setups \cite{zwerger_review}, where locality of noise is crucial for the derivation of error thresholds in this context.
 The fixed point of the EPP is independent of the initial state --provided the fidelity is sufficiently high--, and only depends on noise of the CNOT operations used in the purification. Since the CNOT gates at each step in the EPP only act locally it is plausible that the fixed point
 could be described by local noise as well. For similar reasons it is conjectured in \cite{lnconjecture}
 that the errors introduced by the purification procedure are approximately local.
 Here we systematically investigate if this is indeed the case.

\subsection{General procedure \label{sec:investigate} }
 We start by describing the general procedure we use to perform this investigation.
 First the entanglement purification protocol with imperfect CNOTs is applied until the fixed point $\rho_\mathrm{fp}$
 is reached. The imperfections are considered to
 come from the noisy CNOT operations used in the protocol. This also assumes that the initial state has
 a sufficiently high fidelity so that the protocol can be successful for a given error parameter. Notice that the specific form of the initial state is irrelevant, as the fixed point of the EPP solely depends on noise in the CNOT gates.

In order to see if the resulting state can be described by a local noise model, we proceed as follows.
 The resulting state is compared to one that is obtained by applying local Pauli-diagonal
 noise channels on the desired graph state:
 \begin{equation}
  \rho_L = \mathcal{E}^1(\vec{p}_1) \mathcal{E}^2(\vec{p}_2) \dots \mathcal{E}^N(\vec{p}_N) \Ketbra{G}{G} \text{.}
  \end{equation}
  Now we optimize these noise parameters $\vec{p}_i$ in order to maximize the fidelity with respect to
  $\rho_L$ and $\rho_\mathrm{fp}$
  \begin{equation}
   \mathcal{F}(\rho_L,\rho_\mathrm{fp})= \mathrm{tr} \sqrt{\sqrt{\rho_L} \rho_\mathrm{fp} \sqrt{\rho_L}} \text{.}
  \end{equation}
  If $\mathcal{F}=1$ can be reached, the noise at the fixed point of the EPP can be described
  by a local noise model. Since $1-\mathcal{F}$ is a distance measure \cite{niechu}
  it describes the distance to the closest locally noisy state.
  
  The number of coefficients to be optimized can be reduced by using symmetries and instead of optimizing all coefficients at the same time, only optimize the ones belonging to one set of qubits while keeping the others fixed. The optimization itself is done using an algorithm for sequential least square programming.

\subsection{GHZ states \label{sec:ghzstates}}
We start by considering GHZ states that are local unitary equivalent to graph states where one particle is connected to $N-1$ others, see figure \ref{fig:ghz4}. This can easily be checked by using equation \eqref{eqn:graphstate} for this particular graph. GHZ states are of particular interest since they are resource states for a measurement-based implementation of encoding, decoding and error syndrome readout for a repetition code that can correct bit-flip or phase-flip errors.

  \subsubsection{Binary-like mixtures \label{sec:binlikemix_ghz}}
  We first consider the graph state version of the GHZ state with $N$ qubits (figure \ref{fig:ghz4})
  and a simplified noise model.
  The binary-like mixture is a toy model that describes a hypothetical
  situation for two-colorable graph states in which one color (all qubits in set $A$) is only affected by $\sigma_x$
  noise while the other color (all qubits in set $B$) is only affected by $\sigma_z$ noise \cite{adbepp}. Notice that the different kinds for noise for sets $A$ and $B$ are a consequence of using the graph state formalism. If one considers GHZ states in the standard basis, this corresponds to $\sigma_x$-noise acting on all particles.
  Using the graph state formalism, the density operator can then be written as
  \begin{equation}
   \rho = \sum_{\bm{\mu}_B} \lambda_{\bm{\mu}_B} \Sketbra{\bm{0},\bm{\mu}_B}{\bm{0},\bm{\mu}_B}{G} \text{.}
  \end{equation}
  The CNOT operations used in the EPP are adjusted to fit this model by choosing either $\mathcal{D}_x$ or
  $\mathcal{D}_z$ as the local error channel in \eqref{eqn:mcnot} as appropriate. Note that only one
  subprotocol is necessary for this special model and its effect is given by simply squaring
  the coefficients and then renormalizing the density matrix.

  Using the rules for Pauli-noise on graph states \cite{graphstates} it is easy to see
  that with $\sigma_x$ noise on set $B$ and $\sigma_z$ noise on set $A$
  as well as only $\sigma_x$ noise on set $A$ can always be described
  by a local noise model. In all these cases, one can always find Pauli noise channels acting on the graph state that describe the corresponding density operators.

\begin{figure}[ht]
  \includegraphics[width=\columnwidth]{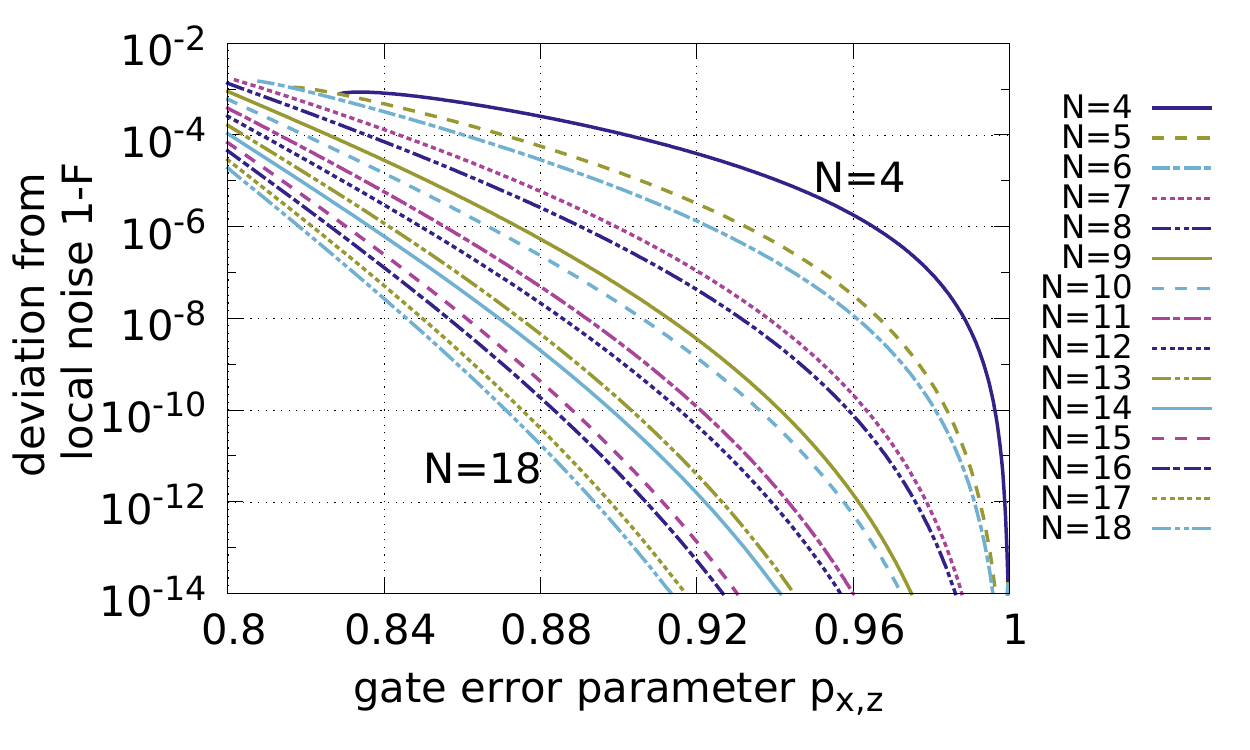}
  \caption{\label{fig:ghzbinarylike} Deviation from a local noise model as a function of the gate error
	  parameter $p_{x,z}$ for the binary-like mixture GHZ-state with only $\sigma_x$ noise acting on the
	  qubit in set $A$ and only $\sigma_z$ noise acting on the qubits in set $B$.}
 \end{figure}

  Less obvious is the case where only $\sigma_z$ noise
  is acting on the qubits in set $B$. However, our numerical simulation shows that this can also be
  described using a local noise model.
  However, for the case where both, $\sigma_x$ errors on set $A$ as well as $\sigma_z$ errors
  on set $B$, occur, we find that for the resulting state at the fixed point of the EPP, this
  is no longer the case for $N>3$. In figure \ref{fig:ghzbinarylike} this deviation from the
  closest local noise model is shown.
  We can easily obtain the results even for high $N$ as it
  is sufficient to consider $\mathcal{O}(N)$ coefficients for the binary-like mixture (see Appendix \ref{sec:On}). This is due to the permutation symmetry of the problem among all qubits in set $B$, which allows one to reduce the number of parameters to describe the density operator from exponentially to lineary many.

\begin{figure}[ht]
  \includegraphics[width=\columnwidth]{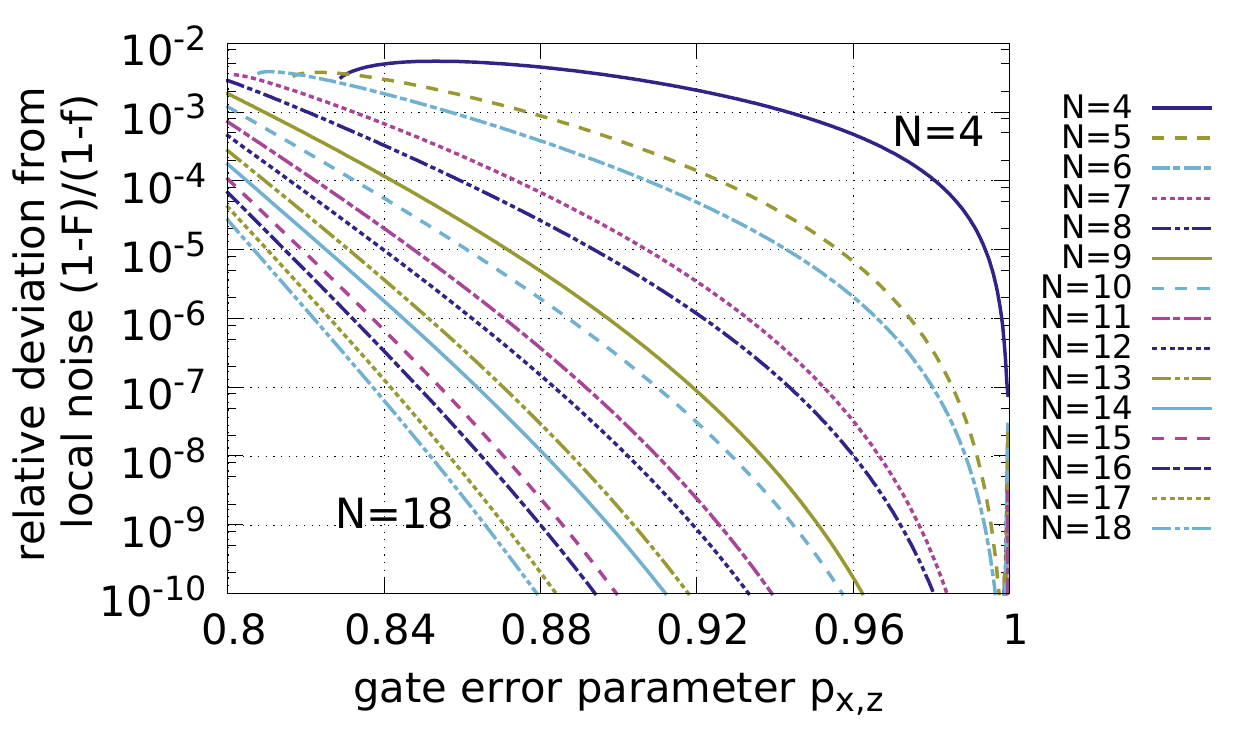}
  \caption{\label{fig:ghzbinarylike_rel} Relative deviation from a local noise model as a function of the gate error
	  parameter $p_{x,z}$ for the binary-like mixture GHZ-state with only $\sigma_x$ noise acting on the
	  qubit in set $A$ and only $\sigma_z$ noise acting on the qubits in set $B$.}
 \end{figure}

  While it is interesting to note that the absolute deviation $1-\mathcal{F}$ is small, a more meaningful measure is
  the relative deviation $(1-\mathcal{F})/(1-f)$ where $1-f$ is the distance between the fixed point state and the perfect
  resource state.
  It measures how big of a mistake we would make in the worst case if we were to replace the fixed point state by the closest
  locally noisy state compared to the error we get by not having the perfect state available.
  This measure can be understood
  as an estimate of the fraction of noise that cannot be attributed to local noise processes.
  The according results in figure \ref{fig:ghzbinarylike_rel} clearly demonstrate that the deviation from a local noise model is very small.

\begin{figure}
  \includegraphics[width=\columnwidth]{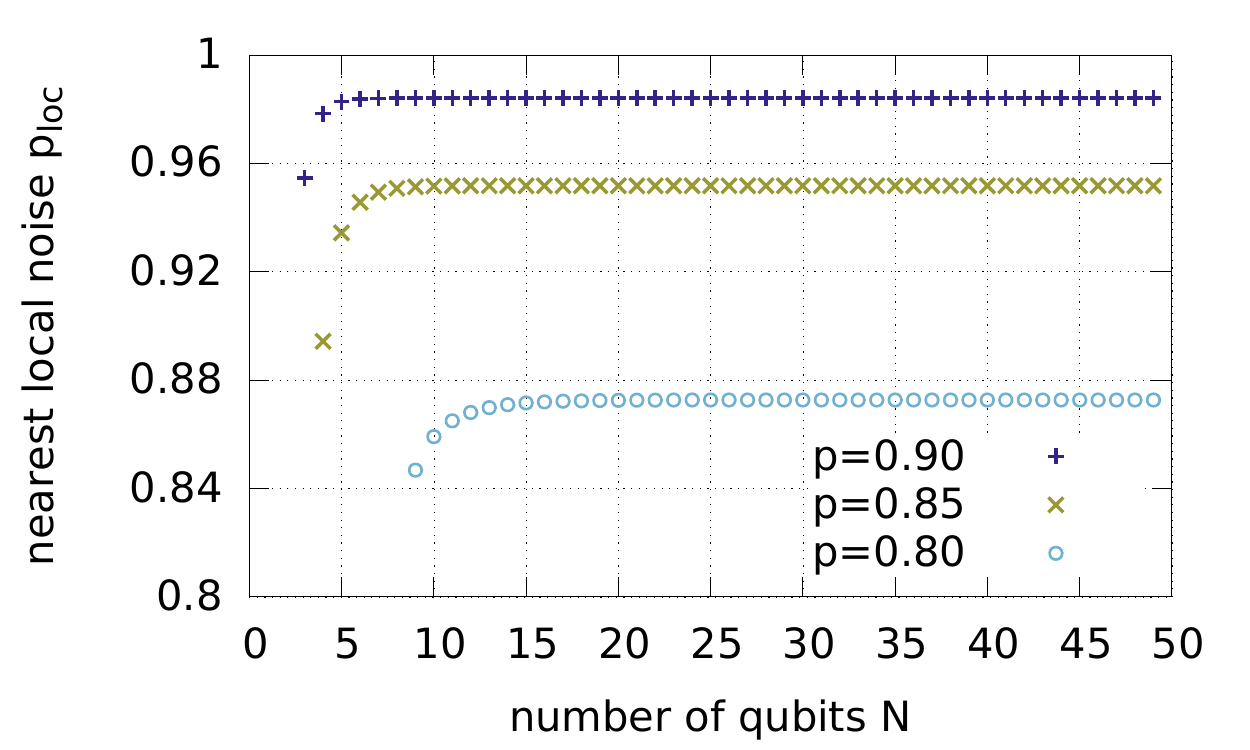}
  \caption{\label{fig:ghzbinary_locnoise} The noise per qubit for the local noise model closest to the fixed point state of the EPP for the binary-like mixtures GHZ-states of varying size.}
 \end{figure}

 This restricted noise model does not only lend itself particularly well to analyse for a high number of qubits because of the reduced number
 of coefficients, but also because the threshold for the gates used for the EPP does not get increasingly restrictive with the size of the GHZ state.
 In figure \ref{fig:ghzbinary_locnoise} the error parameter for the local noise model closest to the fixed point of the EPP is shown. This indicates that
 the noise per qubit stays constant for high number of qubits.
 
 \subsubsection{Depolarizing noise}

\begin{figure}[ht]
  \includegraphics[width=\columnwidth]{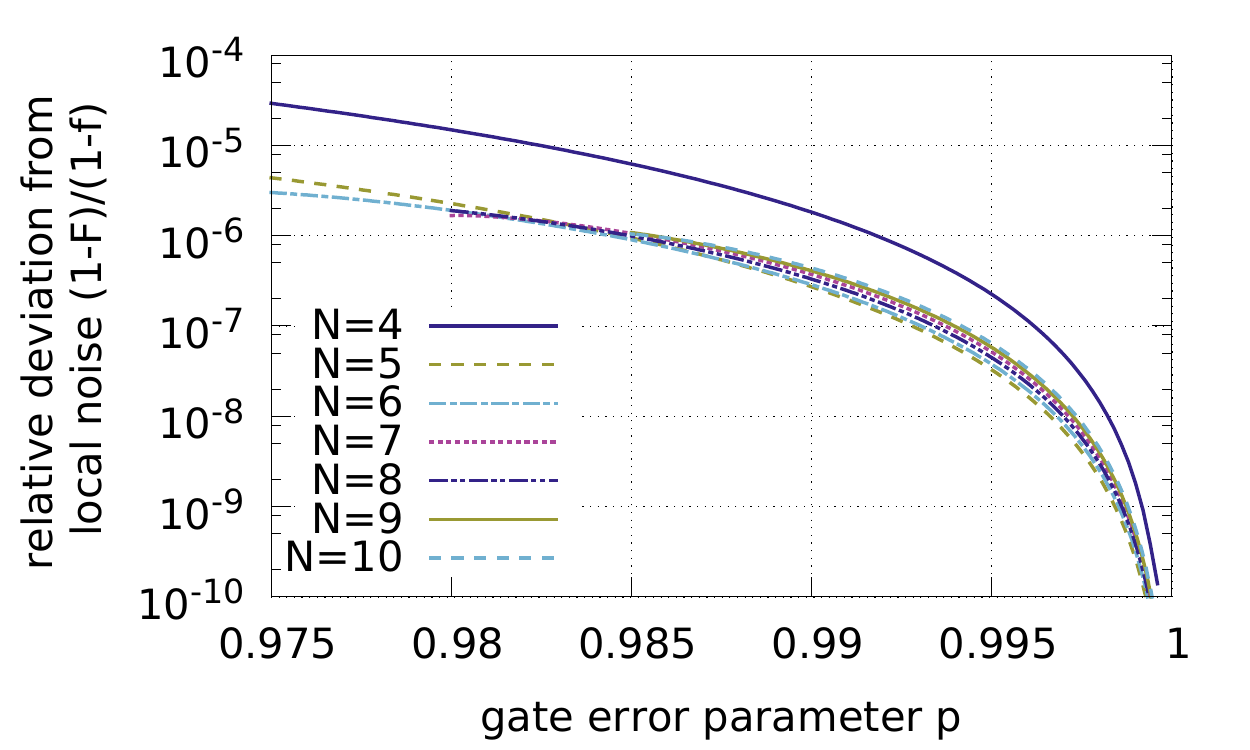}
    \caption{\label{fig:ghz_rel} Relative deviation from a local noise model at the fixed point of the multipartite
	     EPP for GHZ-states with $N\geq4$ qubits. The gate error parameter $p$ refers to the local depolarizing noise used as
	     the noise model for the CNOT gates applied in the EPP.}
 \end{figure}

 Next we consider the situation of general noise. Even though the error correction code is not suited to deal with this kind of noise, as only bit-flip or phase-flip errors can be corrected, it is nevertheless interesting to consider such a situation to assess the performance of the measurement-based approach. We take a look at local depolarizing noise as the noise model of the imperfect CNOT operations
 in the EPP. Again we look at the GHZ state of various sizes. Similar to the simplified model
 a local noise model is not able to describe the state at the fixed point of the EPP (see figure \ref{fig:ghz_rel}).

 We find that the deviation is rather small for both of these cases. This implies that a local noise model is a good
 approximation if these resource states are generated by gate-based entanglement purification, however the noise is not exactly local but in general also contains some non-local component.

\subsubsection{Alternative noise model for noisy gates}

\begin{figure}[ht]
  \includegraphics[width=\columnwidth]{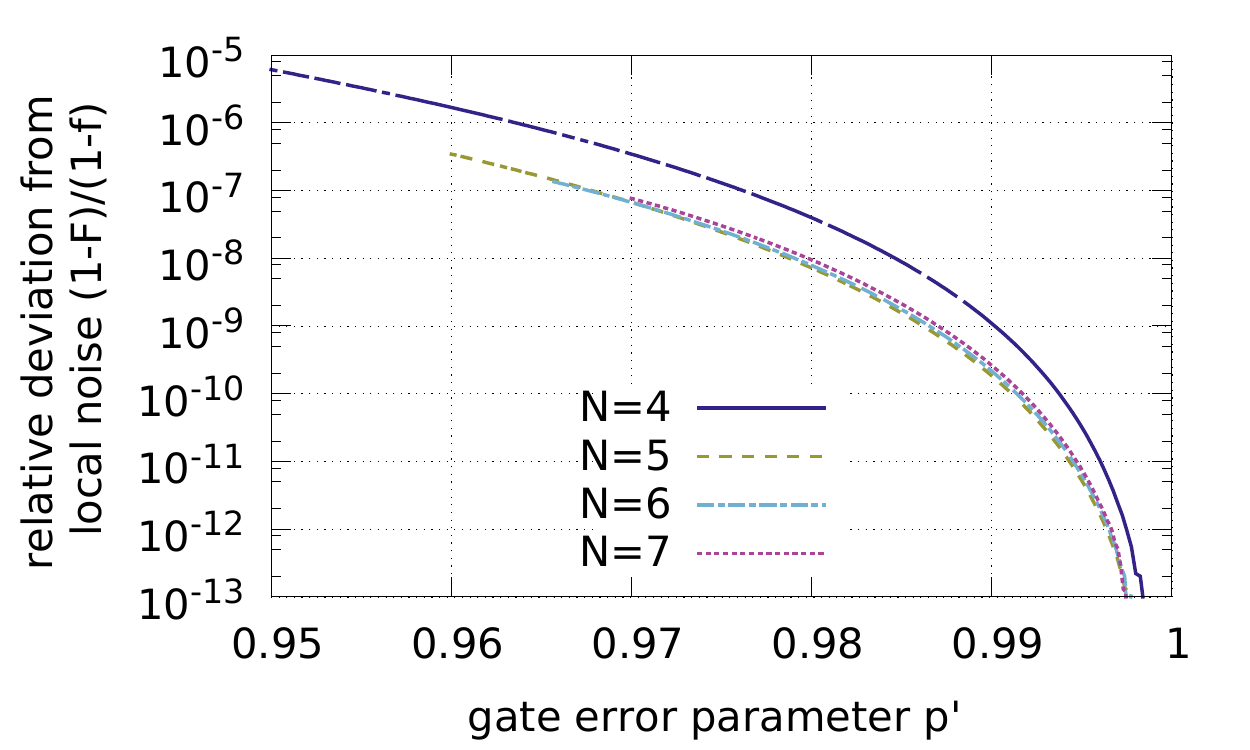}
    \caption{\label{fig:ghz_nonlocalcnot} Relative deviation from a local noise model at the fixed point of the multipartite
	     EPP for GHZ-states with $N\geq4$ qubits. The gate error parameter $p'$ refers to the two-qubit depolarizing noise used as
	     the noise model for the CNOT gates applied in the EPP.}
 \end{figure}

 To verify that the description by a local noise model of the states resulting from entanglement purification is a feature of entanglement purification as such, and not of the particular (local) error model used to describe noisy gates, we have also investigated different error models for CNOT operations. We considered a model where noise in the two-qubit CNOT operation is fully correlated, which we describe by two-qubit depolarizing noise with error parameter $p'$ that acts on the two qubits prior to the application of a perfect CNOT operation. In figure \ref{fig:ghz_nonlocalcnot}
 the results for this model is depicted and we see that the deviation also remains small in this case. However, it is much more challenging to investigate states with a higher number of qubits in this case.
 Because of the correlated noise pattern we need to treat the whole $2^{2N}$-dimensional system instead of the tensor product of two copies with $N$ qubits.

\subsubsection{\label{sec:localize} Localizing noise for GHZ states}
For GHZ states it is possible to bring the state to a form that can be exactly described by local noise. This is done by further manipulating the resulting state after the EPP by means of stochastic local operations, thereby making it compatible with a local error model at the cost
of reducing the fidelity of the state slightly.

We make use of a standard form for GHZ states \cite{Du99, Du00} which can be written using the graph state basis as
\begin{equation}
 \begin{aligned}
  \lambda_{\bm{0}}^+ \ketbra{\Psi^+_{\bm{0}} }{\Psi^+_{\bm{0}} } + \lambda_{\bm{0}}^- \ketbra{\Psi^-_{\bm{0}} }{\Psi^-_{\bm{0}} } + \\ + \sum_{ \bm{k} \neq \bm{0} } \lambda_{\bm{k}} \left( \ketbra{\Psi^+_{\bm{k}} }{\Psi^+_{\bm{k}} } + \ketbra{\Psi^-_{\bm{k}} }{\Psi^-_{\bm{k}} } \right)
 \end{aligned}
 \label{eqn:ghzstandardform}
\end{equation}
 with $\ket{\Psi^+_{\bm{0}} } = \ket{G}$, $\bm{k}=(k_2,k_3,\dots,k_n)$ and $\ket{\Psi^+_{\bm{k}} } = \ket{0,\bm{k}}_G$ as well as
 $\ket{\Psi^-_{\bm{k}} } = \ket{1,\bm{k}}_G$. Every state can be brought to this form without changing the fidelity $\lambda_{\bm{0}}$ by the probabilistic application of local Clifford operations (see \cite{Du00} and Appendix \ref{sec:localizing_details} ).

 The key ingredient in the localization scheme is that $\rho_{\bm{k}} = \left( \ketbra{\Psi^+_{\bm{k}} }{\Psi^+_{\bm{k}} } + \ketbra{\Psi^-_{\bm{k}} }{\Psi^-_{\bm{k}} } \right)$ is a separable
 state that can be easily generated and it is possible to manipulate the coefficients $\lambda_{\bm{k}}$ directly by adding $\rho_{\bm{k}}$ to the ensemble with certain
 weights $q_{\bm{k}}$: $\rho \mapsto Q \rho + \sum_{\bm{k}} q_{\bm{k}} \rho_{\bm{k}}$ with $Q = 1 - \sum q_{\bm{k}}$. We can choose the $Q$ and $q_{\bm{k}}$ in such a way that the
 resulting state is compatible with a local noise model and do so in a way that the fidelity $F$ is decreased as little as possible (see Appendix \ref{sec:localizing_details} for full protocol). We only remark here that the choice of Pauli noise channel with equal $\sigma_x$ and $\sigma_y$ noise in set $A$, and equal $\sigma_y$ and $\sigma_z$ noise in set $B$ leads to states of above standard form, with certain restrictions on the $\lambda_{\bm k}$. One hence only needs to adjust the coefficients $\lambda_{\bm k}$ in such a way that they are compatible with such a local noise process, which can always be done.

\begin{figure}[ht]
 \centering
 \subfloat[\centering \label{fig:ghz_more_local4}]{ \includegraphics[width=0.48\columnwidth]{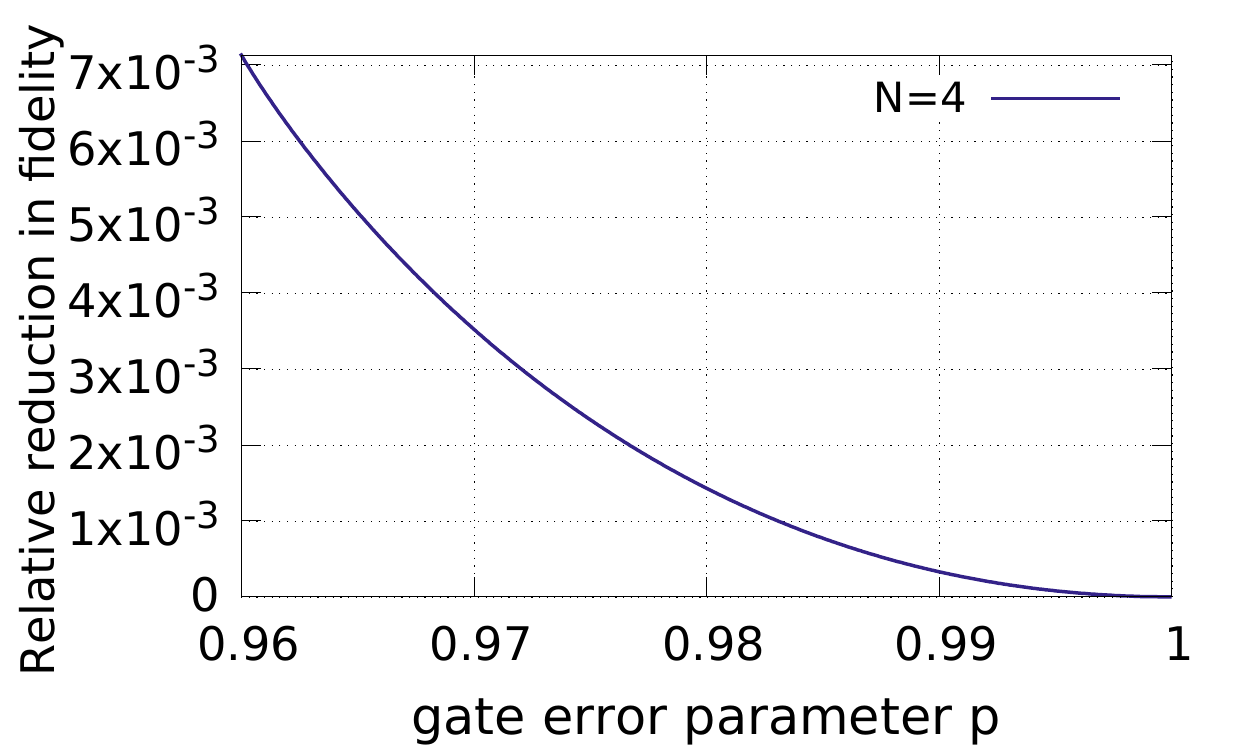}  }
 \subfloat[\centering \label{fig:ghz_more_local9}]{ \includegraphics[width=0.48\columnwidth]{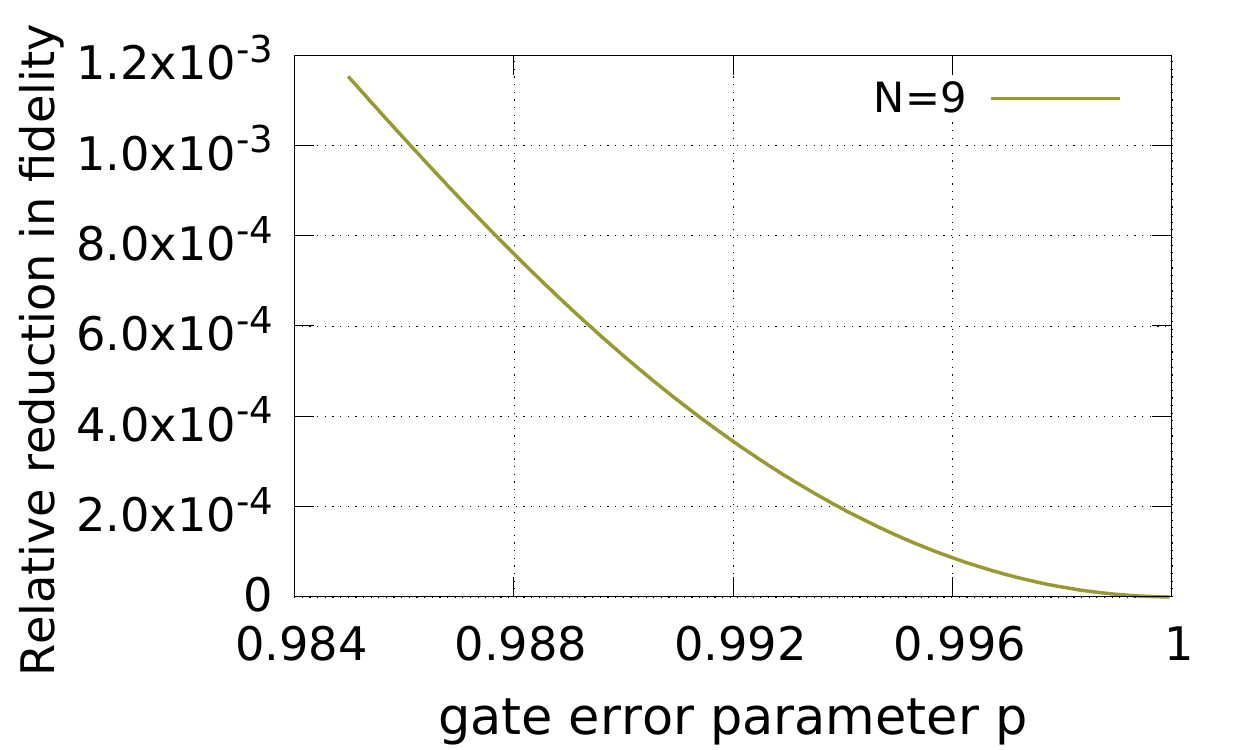}  }
 \caption{\label{fig:ghz_more_local} The relative reduction in fidelity $(F-F')/F$ for applying the noise localisation procedure at the fixed point of the multipartite EPP
 with local white noise as the error model for the  CNOT gates for
  (a) $N=4$ (b) $N=9$ qubit GHZ states relative to the fidelity of the fixed point state.}
\end{figure}

 In figure \ref{fig:ghz_more_local} examples for the relative reduction in fidelity $(F-F')/F$ of this localization procedure for GHZ states
 that had been purified using noisy gates with a white noise error model are shown. One sees that since the deviation from a local noise description is small to start with, the localization procedure for noise results in only a very small reduction of the fidelity. Hence noisy GHZ states resulting from EPP can always be brought to a form that they can be described by a local noise model.

\subsection{Cluster-ring code}
\begin{figure}[ht]
 \includegraphics[width=0.3\columnwidth]{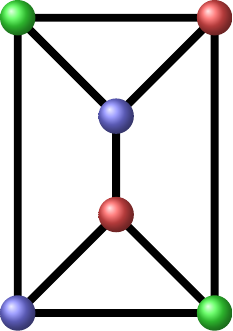}
 \caption{\label{fig:lc_clr} The three-colorable 6-qubit graph state that is local unitary equivalent to the resource state implementing
	  the cluster-ring code.}
\end{figure}

We now consider the cluster-ring code for measurement based error correction and
study the corresponding resource state (figure \ref{fig:clusterringcode}) for encoding and decoding. This code is capable of correcting one arbitrary error occurring on one of the qubits, and we will now consider depolarizing noise.
For the EPP it is beneficial to consider a different graph state (see figure \ref{fig:lc_clr}), which is local unitary
equivalent to the desired state and can be obtained by applying the local complementation rule \cite{graphstates} on the red qubit in the middle in figure \ref{fig:lc_clr}.
This state has the advantage that it is three-colorable rather than four-colorable, which makes it more efficient to purify as fewer different subprotocols are needed (see Appendix \ref{sec:purdetails}).

\begin{figure}[ht]
 \centering
 \includegraphics[width=\columnwidth]{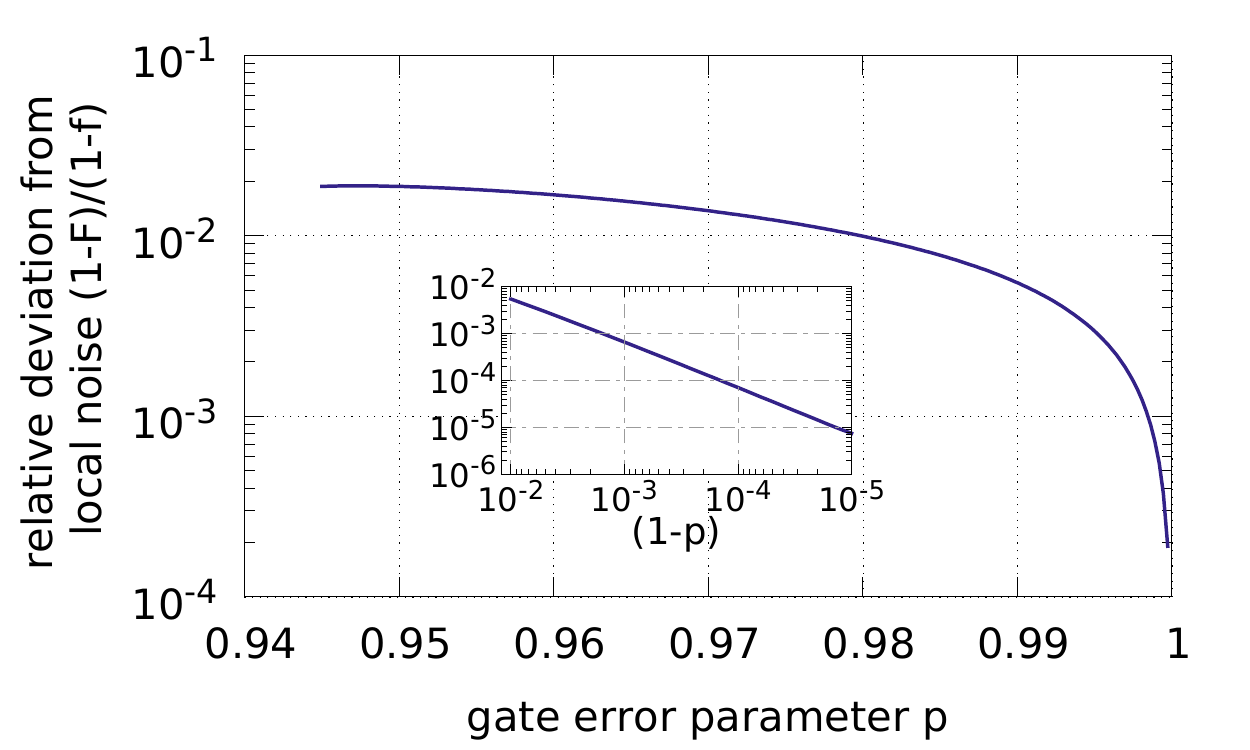}
 \caption{\label{fig:noisestruct_clr5} Relative deviation from a local noise model for the graph state in figure \ref{fig:lc_clr}. The gate
	  error parameter $p$ refers to the local white noise channels of the CNOT gates used in the EPP.}
\end{figure}

The relative deviation from a local noise description of the state resulting after EPP is shown in figure \ref{fig:noisestruct_clr5}. While this relative distance
is not quite as small as for the GHZ state, the error we would make when using this locally noisy state instead of the fixed point state is still nearly two orders of magnitude smaller than the error from not having access to the perfect graph state. In particular, for small noise, i.e. $p$ close to one, the description by a local noise process becomes better.

We have also investigated EPP of other states, e.g. a linear cluster state of five qubits (see Appendix \ref{sec:lcs}). Also in this case we find that a local error model provides a good approximation, though the precision is slightly worse as compared to GHZ states or resource state for cluster-ring code.

\section{Encoded measurement based quantum communication with purified states \label{sec:application}}
The techniques used to obtain high error thresholds for measurement based quantum computation \cite{mbqc_eppthresh,hybrid}
made a key assumption that the imperfect preparation of the resource states results in a state which can be describe as local noise channels on the perfect state.

Here we investigate how the specific noise pattern that arises when preparing resource states using EPPs with noisy gates
influences the performance of measurement-based quantum error correction when using these states as resources instead. In section \ref{sec:noisestruct} we have shown that the local noise model is a good approximation, however, in general the noise cannot be described that way. This difference prevents using the elegant method
used in \cite{hybrid,zwerger_review} for the case of local noise, where noise acting on resource states can simply by shifted to the input particles when performing a Bell measurement. For correlated noise, we can not simply do this \footnote{For correlated noise with correlation only among input particles or output particles, this approach is still possible. However, for correlated noise between input- and output particles, one can no longer use such a simplified analysis.}, but have to do a full analysis of the overall process. We implemented the actual error correction codes numerically for this discussion (See Appendix \ref{sec:dec_patterns} for details.)

\begin{figure}[ht]
 \centering
 \subfloat[\centering]{\includegraphics[width=\columnwidth]{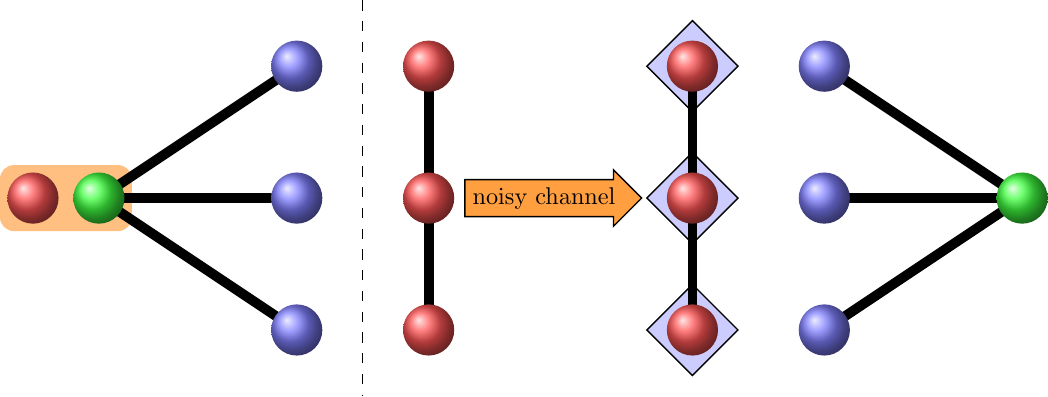} }

 \subfloat[\centering]{\includegraphics[width=0.4\columnwidth]{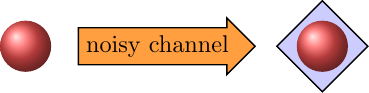} }
 \caption{\label{fig:addextnoise} Illustration of using measurement based quantum error correction in a quantum communication scenario.
	  (a) Scenario (ii): The encoded information is sent through a noisy channel with local noise channels acting on the encoding qubits. Then the
	  information is decoded at the destination using a resource state that has been generated by an EPP. Scenario (iii):
	  Additionally the encoding at the source of the information is also done in a measurement based way using another copy of that resource state (left hand side).
	  These approaches can be compared to (b) sending the information through the noisy channel without encoding.
	  }
\end{figure}

The two error correction codes we analyzed are the measurement based variants of a simple repetition code and the five-qubit cluster-ring code, where the resource states for measurement-based encoding and decoding are shown in figures \ref{fig:ghz4} and \ref{fig:clusterringcode}.
To analyze their performance we use the Jamiolkowski fidelity \cite{jamfid}, which is the fidelity of the Jamiolkowski state Eq. \ref{rho_jam} of the effective map
compared to the ideal case and is also linearly related to the average teleported fidelity \cite{fidnielsen, horo3}.
We consider three different (elements of) communication scenarios (see figure \ref{fig:addextnoise}):
(i) Decoding of a perfect encoded state; (ii) Decoding after the state was transmitted through a noisy quantum channel; (iii) Encoding, transmission through a noisy quantum channel and decoding.

In all three scenarios, we consider a measurement-based implementation of encoding and decoding, where the latter is with built-in error correction.
The resource states are generated by means of noisy EPP. We compare this measurement-based approach with purified states to a direct transmission of the information without encoding, where the effective map for calculating the Jamiolkowski state is simply the external noise channel. We also evaluate the performance of alternative schemes, namely a naive generation of the
graph state by directly applying a CZ operation for each edge in the graph and a completely gate based implementation (Appendix \ref{sec:qeccircuit}) of the decoding procedure.
The error parameters for the CNOT and CZ gates are assumed to be the same for all these approaches. Furthermore, for some cases we also investigate
additional variants of how resource states for the measurement-based implementation are prepared (assuming a certain abstract noise model or different variants of EPP).

\subsection{Decoding only}
We first consider scenario (i) and investigate the errors that the imperfect resource states introduce when they are used to decode a perfect encoded state.
Here we will only show a selection of results that lead to interesting insights about this approach.

\begin{figure}[ht]
 \includegraphics[width=\columnwidth]{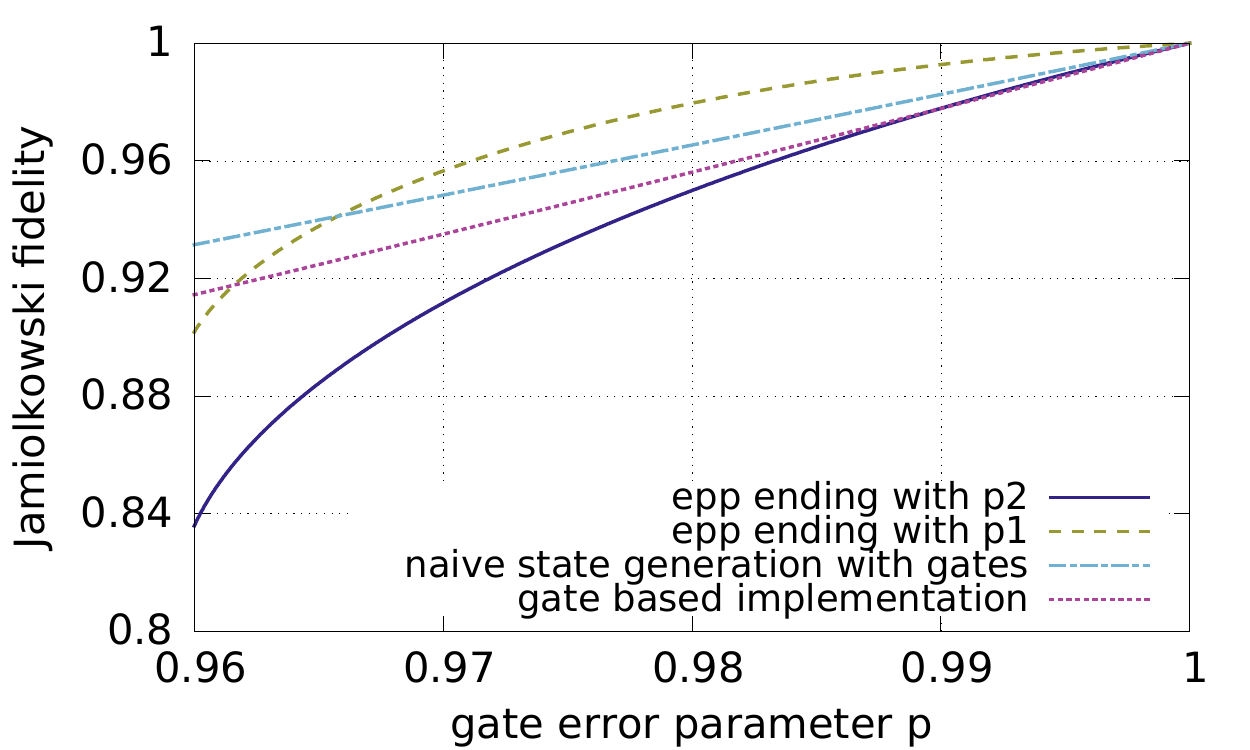}
 \caption{Scenario (i): Implementation of repetition code decoding with resource states generated using noisy CNOT gates with a local white noise
 error model with parameter $p$. Curves correspond to resource states generated by EPP ending with P2 (blue, solid), P1 (green, dashed), or directly via gates and without EPP (light blue, dashed-dotted). This is compared to a gate-based implementation of the decoding procedure using noisy gates (pink, dotted). \label{fig:ghz4_wnoise2} }
\end{figure}

One important discovery is apparent from the results concerning the GHZ state with $N=4$ qubits when the CNOT gates are modelled with
local white noise, a noise model this code is not suited to correct. In figure \ref{fig:ghz4_wnoise2} one can see that whether the alternating EPP is ended with subprotocol P1 or P2 makes a huge difference for the performance of the resource state. In this particular case it can be easily understood that ending with subprotocol P1 is preferable
as it purifies the coefficients belonging to the single qubit in set $A$, which is not protected by the error correction code.
From this we conclude, that the details of the preparation procedure is of central importance to the suitability of the resulting resource state. It is possible that further optimization of preparation (e.g. considering other sequences of P1 and P2 than the alternating one) could improve the performance even further.

\begin{figure}[ht]
 \includegraphics[width=\columnwidth]{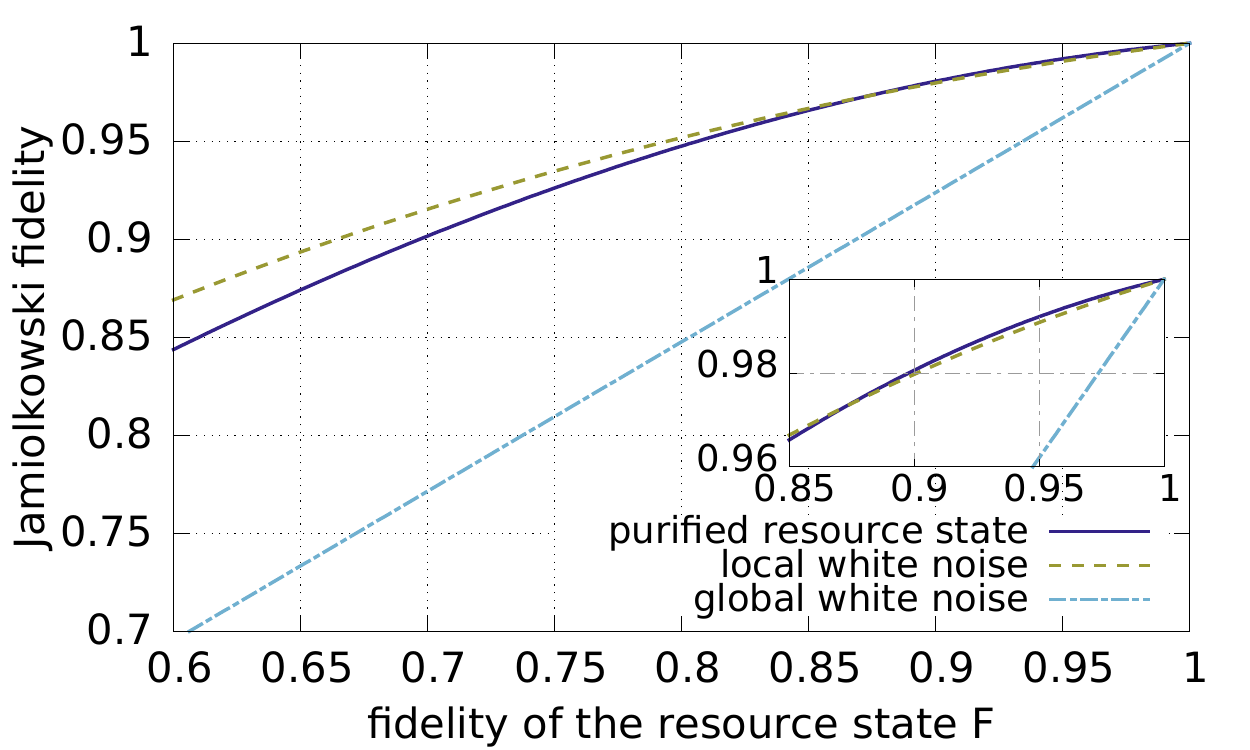}
 \caption{Scenario (i): Comparison of different error models for resource states used for measurement based decoding using the cluster-ring code. The comparison
	  is done by the fidelity of the resulting resource state since the error parameters of the different models cannot be directly compared. \label{fig:clr_byfidelity}}
\end{figure}

Next we analyse the influence of the shape of the noise for resource states on the performance for measurement-based decoding. We compare the performance to the state that is actually generated by entanglement purification with resource states that suffer from global (correlated) or local depolarizing noise.
We compare resource states with the same fidelity to each other, as error parameters for different models are not directly related. The results are shown in figure \ref{fig:clr_byfidelity} for the cluster-ring code. It appears that correlated noise is rather bad for the performance, however EPP produces states that are close to a local noise model. Interestingly, for a range of fidelities the resource state generated by multiparty entanglement purification outperforms the state with a
completely local noise model, if properly optimized (Appendix \ref{sec:purdetails}). Hence, the locality of the noise does not appear to be the key feature when considering measurement based quantum error correction. It is more important that the resulting noise structure can be efficiently corrected by the chosen error correction code and the most suitable state of those that can be obtained with a certain method is not even necessarily the one with the highest fidelity.

\subsection{Encoding and Decoding with external noise}
A more realistic setting in which one has a motivation to make use of error correction is in the presence of external noise, scenario (ii) and (iii).
So unlike the case discussed before, the imperfections in the system do not only arise from imperfect resource state preparation.
We consider a scenario where a purified resource state is used to decode the encoded qubits sent through a noisy channel (scenario (ii), see figure \ref{fig:addextnoise}).
To evaluate this approach we compare it with simply sending the information without encoding, as well as preparing the resource state directly with gates without EPP and a
fully gate-based approach. We also consider the full communication scheme, scenario (iii),
where encoding is also done using the same method. So in a measurement based implementation another copy of the resource state is used.

\subsubsection{\label{sec:ghzvariant_extnoise} Binary-like mixture GHZ state}
First let us consider a purified binary-like mixture GHZ state with an external error which this code is suited to correct, that is a $\sigma_z$ error
channel $\mathcal{D}_z^j (q_z)$ acting on each of the physical qubits that carry the encoded information.

\begin{figure}
 \centering
 \subfloat[\centering \label{fig:extnoise_decode_ghz_a} $p_{x,z}=0.92$]{ \includegraphics[width=0.80\columnwidth]{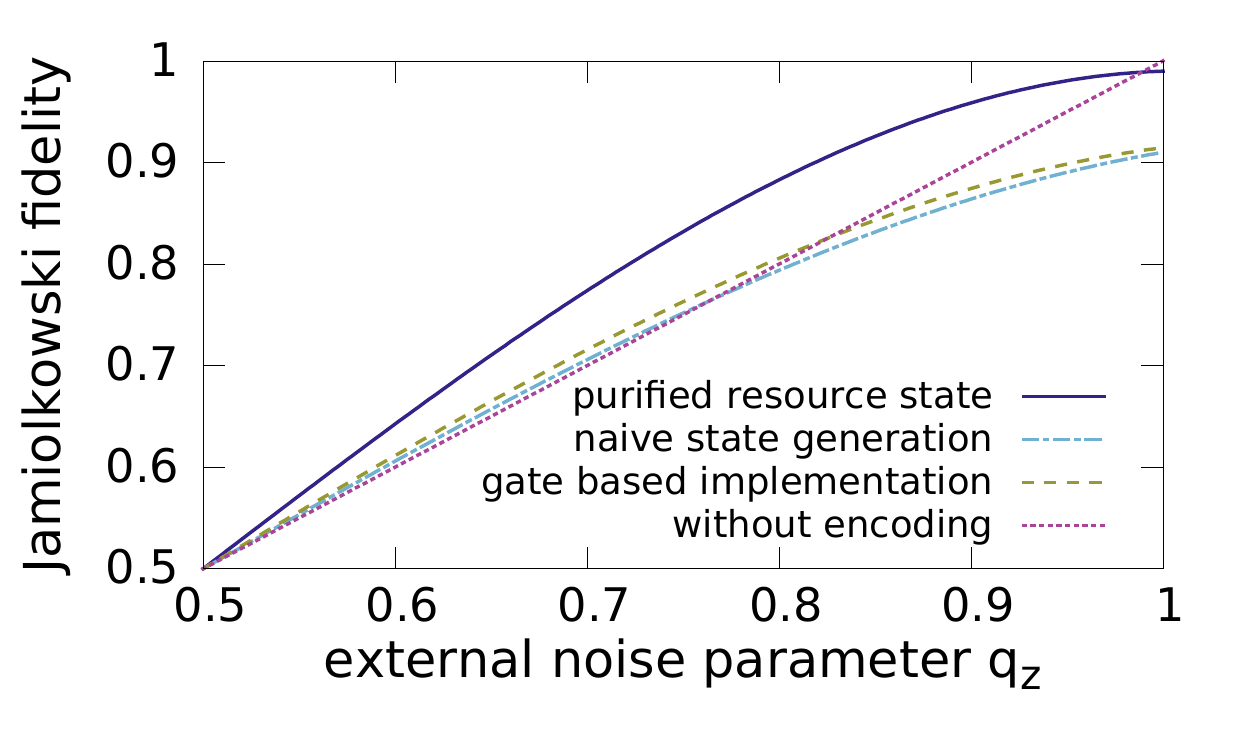}  } \\
 \subfloat[\centering \label{fig:extnoise_decode_ghz_b} $q_{z}=0.90$]{ \includegraphics[width=0.80\columnwidth]{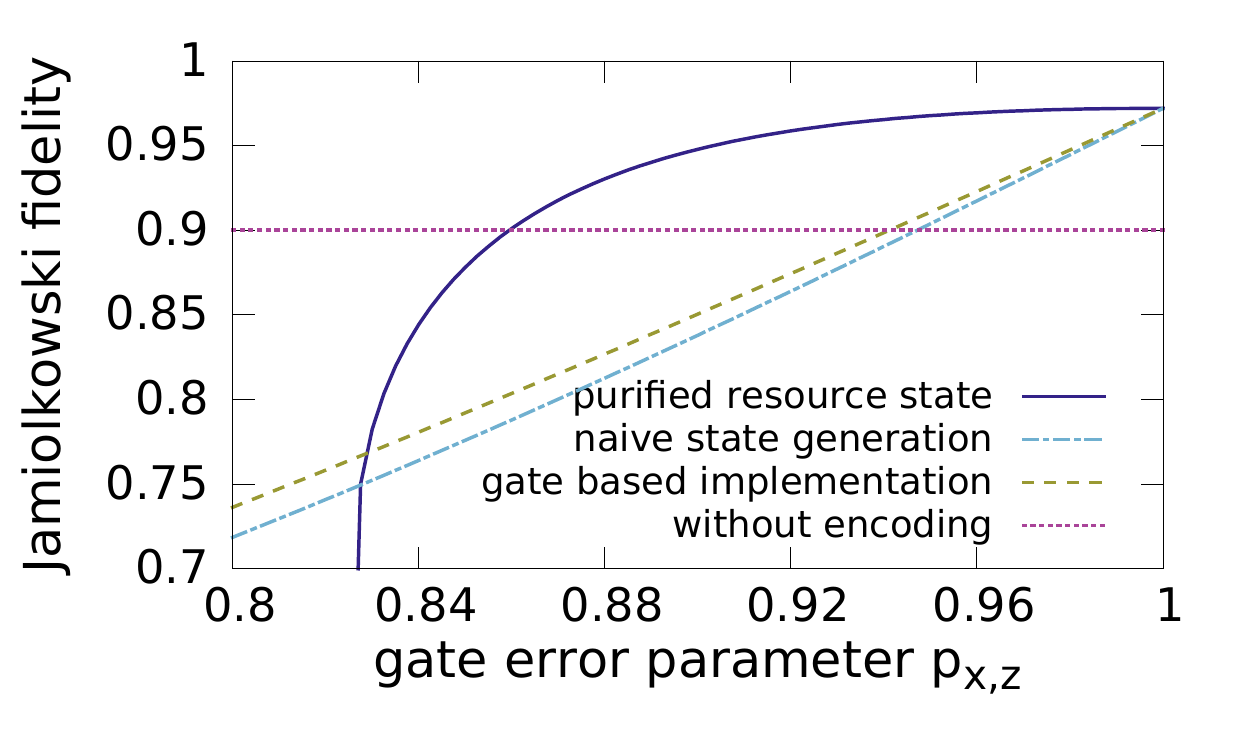}  }
 \caption{\label{fig:extnoise_decode_ghzvariant} Scenario (ii): Comparison of decoding using the purified binary-like mixture GHZ state with
	  a naively generated resource state, a gate based implementation as well as unencoded transmission in the presence of external noise.
	  Results for a fixed valued of the (a) gate error parameter $p_{x,z}$ (b) external noise parameter $q_{z}$.}
\end{figure}

\begin{figure}
 \centering
 \subfloat[\centering \label{fig:extnoise_decode_ghz_pq}]{ \includegraphics[width=0.48\columnwidth]{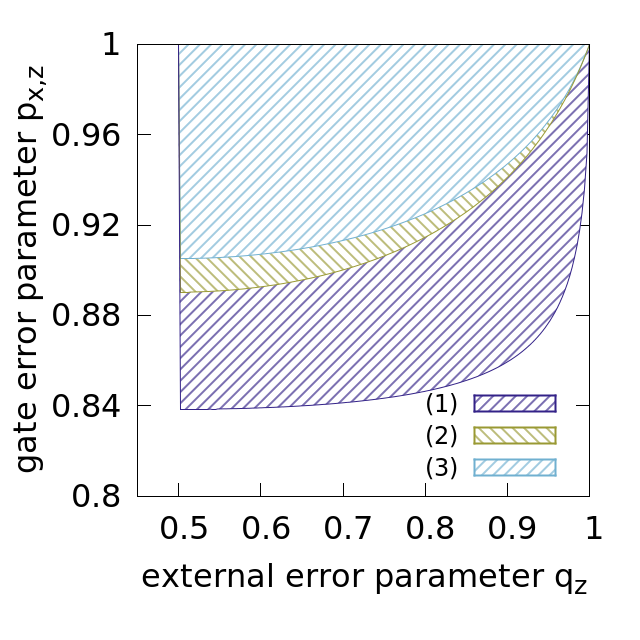}  }
 \subfloat[\centering \label{fig:extnoise_encdec_ghz_pq}]{ \includegraphics[width=0.48\columnwidth]{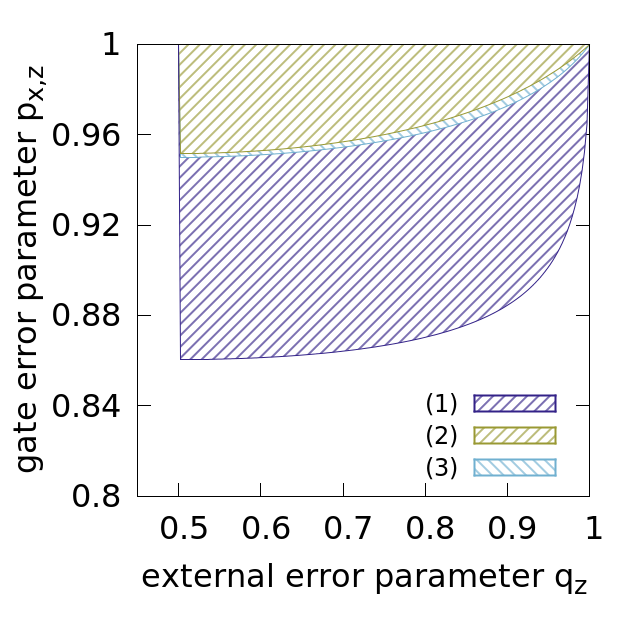}  }
 \caption{Repetition code with a restricted error model. The highlighted areas (and the other areas they enclose) indicate the parameter regimes where
	  the approach using (1) purified resource states; (2) the gate-based implementation; (3) directly generated resource states yields better
	  results than direct transmission of non-encoded information when used for (a) decoding (scenario (ii)) (b) encoding and decoding (scenario (iii)).}
\end{figure}

Figure \ref{fig:extnoise_decode_ghzvariant} shows the results for examples of external noise parameter $q_{z}$ and gate error parameter $p_{x,z}$. When examining that figure
it is apparent that even with noisy resource states this approach can be useful, but if the error rate of the channel is very low ($q_{z}$ close to 1) the imperfect resource state
introduces more errors than it corrects. Furthermore, we see that the measurement-based approach for using states prepared via EPP is superior to a direct preparation of resource states and
to a gate-based implementation.
From these results one can infer that there is a certain range of parameters for which each of these approaches offer an advantage over non-encoded transmission.
These parameter ranges are shown in figure \ref{fig:extnoise_decode_ghz_pq}.
If encoding is no longer done perfectly and both encoding and decoding is performed using the same method (scenario (iii)), the ranges found are
a bit smaller, but using purified resource states is still very useful and superior to the alternatives considered. (see figure \ref{fig:extnoise_encdec_ghz_pq}).

\subsubsection{Cluster-ring code}
\begin{figure}
 \centering
 \subfloat[\centering \label{fig:extnoise_decode_clr_a} $p=0.999$]{ \includegraphics[width=0.80\columnwidth]{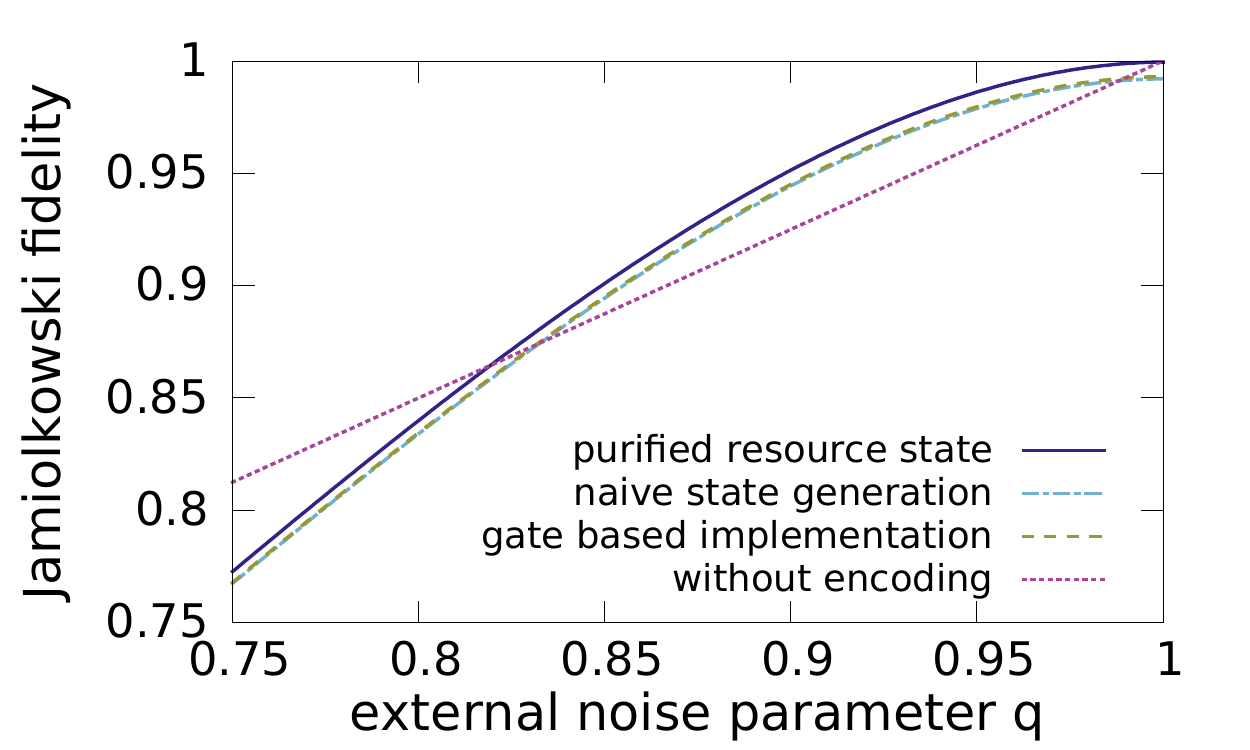}  }\\
 \subfloat[\centering \label{fig:extnoise_decode_clr_b} $q=0.95$]{ \includegraphics[width=0.80\columnwidth]{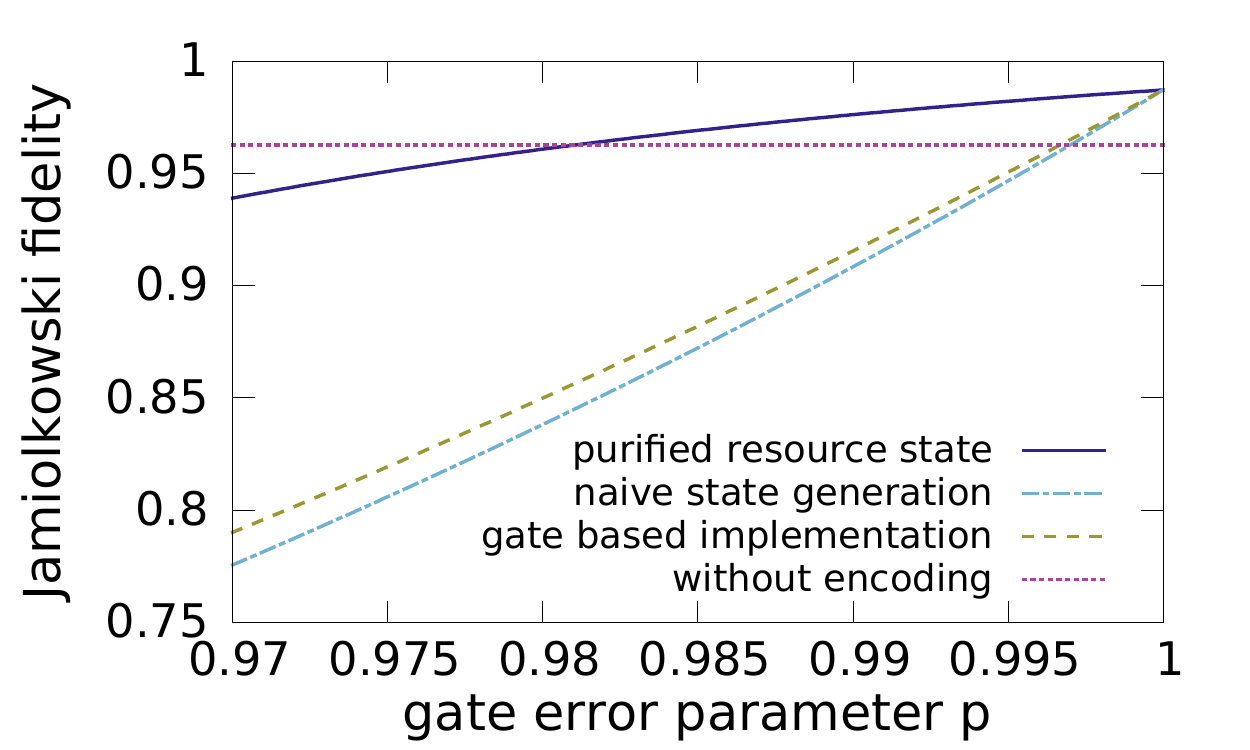}  }
 \caption{\label{fig:extnoise_decode_clr} Scenario (ii):  Comparison of decoding using the purified resource state for the cluster-ring code with
	  a naively generated resource state, a gate based implementation as well as unencoded transmission in the presence of external noise.
	  Results for a fixed valued of the (a) gate error parameter $p$ (b) external noise parameter $q$.}
\end{figure}

We also performed the same comparisons for the cluster-ring code, but with a local white noise channel {$\mathcal{D}_w^j(q)$} as the external noise acting on each encoded qubit.
We consider the gate errors to also be described by local white noise.
The results in figure \ref{fig:extnoise_decode_clr} for some example values look similar to those for the GHZ state with a restricted error. Again, we find that for scenario (ii) the measurement
based approach is beneficial for a wide range of parameters. Using purified resource states again allows for a bigger range than the alternative implementations.

\begin{figure}
 \centering
 \subfloat[\centering \label{fig:extnoise_decode_clr_pq}]{ \includegraphics[width=0.48\columnwidth]{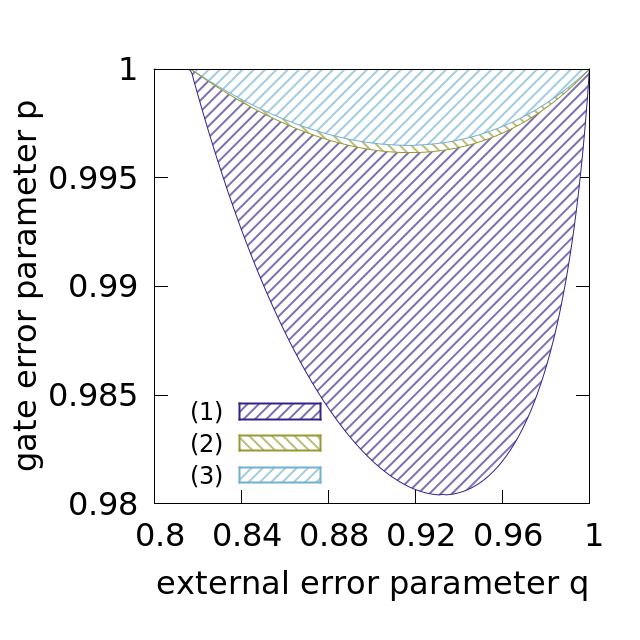}  }
 \subfloat[\centering \label{fig:extnoise_encdec_clr_pq}]{ \includegraphics[width=0.48\columnwidth]{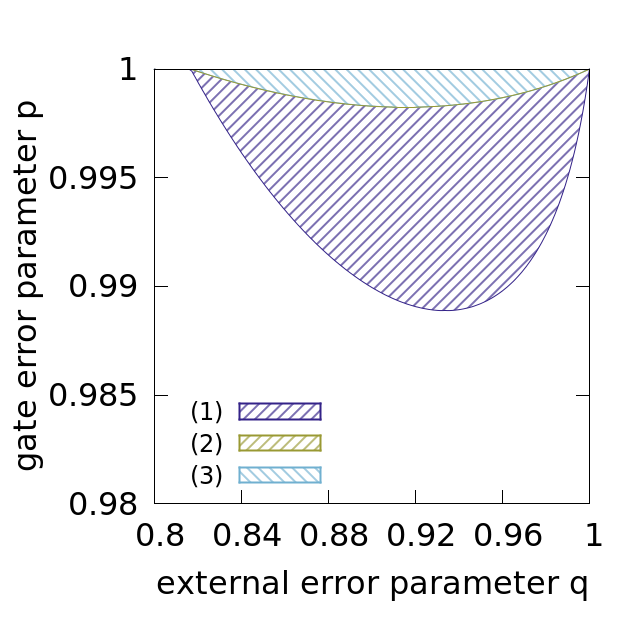}  }
 \caption{\label{fig:extnoise_clr_pq} Cluster-ring code with local white noise. The highlighted areas (and the other areas they enclose) indicate the parameter regimes where
	  the approach using (1) purified resource states; (2) the gate-based implementation; (3) directly generated resource states yields better
	  results than direct transmission of non-encoded information when used for (a) decoding (scenario (ii)) (b) encoding and decoding (scenario (iii)).}
\end{figure}

In figure \ref{fig:extnoise_clr_pq} the parameter ranges for which each of the approaches remains useful are shown for scenarios (ii) and (iii). The results
look very similar but lack the sudden cut-off at a specific external noise level observed for the restricted noise model. 
Most importantly, we observe that using the purified resource states allows to use error correction in parameter regimes where the other approaches fail.

\section{\label{sec:scaling} Fault-tolerance and scaling for a restricted error model}
While very high error thresholds have been found for measurement based setups, these thresholds cannot be directly compared to previous results concerning fault-tolerance in a gate-based approach because the underlying error models are completely different. In the measurement based case one usually considers an abstract error model where local noise is acting on the perfect resource state while for the gate based implementation the error parameters specifically describe the individual gates used in the circuit. The approach of using states generated by gate based entanglement purification in measurement based quantum computation tasks could be a promising avenue to find a relation between these two computational models.

The hybrid model of quantum computation \cite{hybrid} is a natural starting point for this investigation as it uses relatively small-scale resource states. In the hybrid model one makes use of the modular nature found in the circuit model (only single-qubit operations and an entangling gate are needed for universal quantum computation), but one implements these elementary building blocks in a measurement based way. However, even for this manageable class of states the analysis is not straightforward. The results presented here should be understood as a first step towards bridging the gap between different error models for different quantum computation architectures and we would like to point out some of the challenges involved in this process.

For a full analysis of fault tolerance in the hybrid model, one would need to investigate resource states for all encoded Clifford gates. Furthermore, it would also require the implementation of non-Clifford gates, e.g. via code switchers and transversal gates or via magic state distillation \cite{magicstatedist}. Here we restrict ourselves to one qubit Clifford operations that are implemented in an encoded way, so in particular we are interested in resource states that implement decoding, error detection and encoding at once as these are of central importance to determine the thresholds for the hybrid approach.

However, to actually obtain the fault-tolerance thresholds we would have to consider concatenated error correction codes and their scaling with an increasing number of encoding qubits, which poses obvious problems for a numeric analysis. We thus restrict ourselves even further to a scenario with an error model, where only bit-flip errors occur. For this type of noise it is sufficient to use the simple bit-flip code for error correction. Because both the resource states for encoding and decoding are given by $N+1$ qubit GHZ states one obtains that the combined resource state for decoding and encoding with built-in error correction is simply a $2N$ qubit GHZ state with $N$ input and $N$ output qubits (\cite{hybrid} describes how such resource states are constructed in general).

This again leads to the toy model of binary-like mixture GHZ states, where we can obtain results for a large number of qubits and investigate the scaling behavior of the underlying error correction code. We consider a situation where the resource state is generated using an EPP and is then purified to the fixed point of the protocol. The error threshold for which it is still beneficial to use a measurement-based implementation of this error correction code is determined by the smaller one of the following two thresholds: (i) Threshold for EPP, i.e. noise of CNOT gates such that EPP for producing the required resource states (in this case GHZ states of size $2N$) works; (ii) Threshold for measurement-based error correction itself, i.e. of the resulting resource state such that error correction leads to a smaller, in the asymptotic limit vanishing noise at the logical level.

Both thresholds are determined by the gate noise: In case of (i), the acceptable gate noise directly provides the error threshold for the EPP. In case of (ii), the noisy gates determine the achievable fixedpoint of the EPP, and hence the noise of the resource state. This noise, in turn, determines if a successful error correction is possible \footnote{Notice that we assume perfect Bell measurements here, as discussed in Sec. \ref{sec:errormodel}.}.
Regarding (i), error thresholds for noisy EPP have been studied in detail in \cite{adbepp,eppallgraphs,eppreview}. These results already provide a bound 
for the applicability of the approach. For the case studied here if the entanglement purification fails, 
the error correction also proves to be insufficient, so we concentrate on (ii) in the following.

\begin{figure}
  \includegraphics[width=\columnwidth]{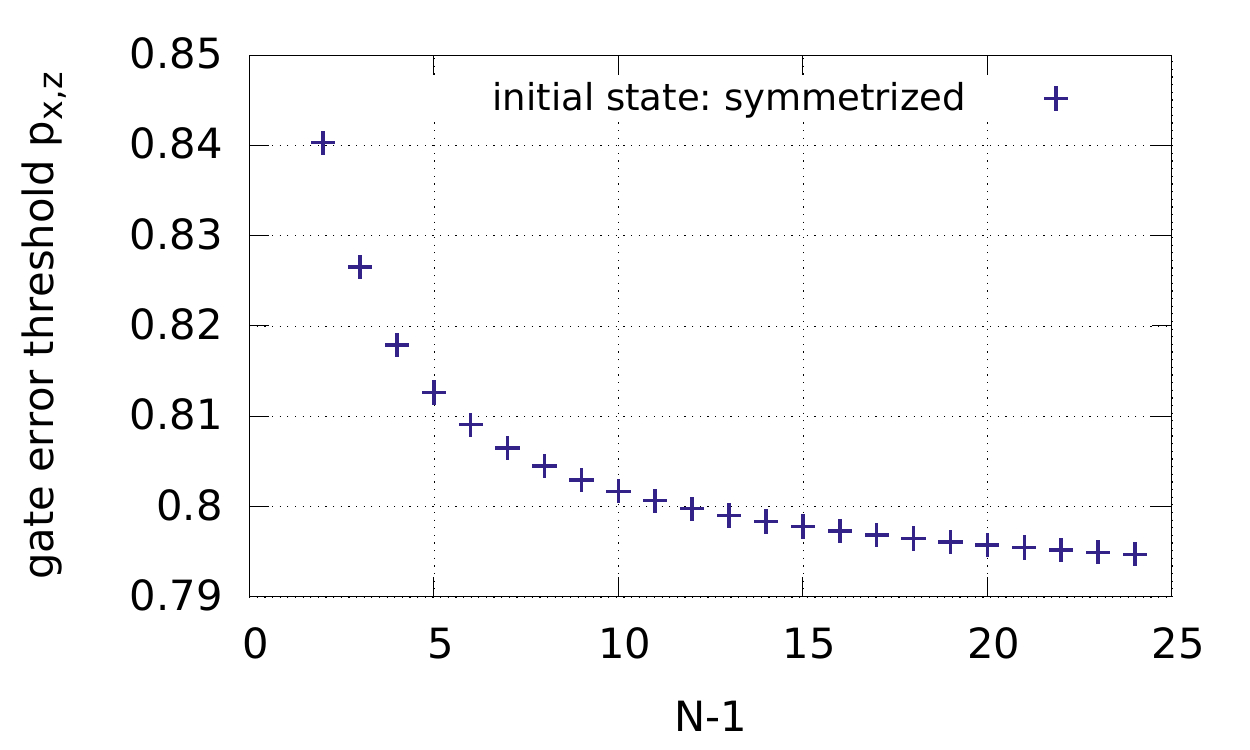}
  \caption{\label{fig:prep_thresh} Threshold for the gate error parameter $p_{x,z}$ of the entangling gates such that the initial state generated using
	    these gates can still be succesfully purified with the EPP using the same gates.}
\end{figure}

In addition, the initial states which are used as inputs for the EPP need to be prepared. If one would do this via entangling gates, one needs to make sure that the resulting states are distillable by the noisy EPP. While this limit is no issue for the analysis in the previous sections, it is very relevant when investigating the scaling behavior with many qubits. We generate the initial $N$-qubit GHZ states with the gate sequence 
\begin{equation}
  U_\mathrm{CNOT}^{(N-1) \rightarrow N} \dots U_\mathrm{CNOT}^{2\rightarrow 3} U_\mathrm{CNOT}^{1\rightarrow 2} \left(\Ket{0}+\Ket{1}\right)/\sqrt{2} \otimes \Ket{0}^{\otimes N-1} \text{.}
\end{equation}
and then completely symmetrise the resulting state \footnote{The symmetrisation can be done locally by randomly reenumerating the qubits.}. When considering that for the binary-like mixture the effect of the EPP is simply squaring the coefficients and renormalizing it is apparent that the second highest coefficient is the limiting factor for whether a state is still distillable. The symmetrisation is essential to obtain a good threshold as it equalises multiple coefficients without changing the fidelity and therefore reduces the second highest coefficient.
In figure \ref{fig:prep_thresh} the threshold for the gate noise such that the state prepared this way is still distillable is shown. While this threshold is quite restrictive,
keep in mind that for certain physical setups there exist better ways to create a GHZ state such as the Mølmer-Sørensen gate \cite{molmersorensen} for trapped ions.
Hence we continue our analysis below by assuming that the the fixed point states of the EPP for a given gate error parameter $p_{x,z}$ can be reached and do not explicitly take the initial state preparation
into account.

However, even with all these restrictions placed on the scenario there are still some obstacles to obtaining the threshold value. In order to relate the error threshold of our approach to the known fault-tolerance threshold of the error correction code obtained by using the techniques of \cite{hybrid}, the noise on the resource state needs to be local so it can be moved from the input qubits of the resource state to the input state. As shown in section \ref{sec:binlikemix_ghz} the state at the fixed point of the EPP cannot be described by a local noise model and therefore even for the simple noise model in the binary-like mixture case, the threshold is not easily analysed. It should be noted that the protocol to localize the noise for GHZ states in section \ref{sec:localize} is not applicable here, as we would leave the subspace in which the binary-like mixture states are located (and thus introduce errors the code is unable to correct). We take a look several different approximations in order to estimate the threshold for the noise of the CNOT gates used in the EPP:

\begin{itemize}
 \item \textit{Scenario A:} A state with the closest local noise model found by the optimization procedure described in section \ref{sec:investigate} is considered instead of
the fixed point state.

\item \textit{Scenario B:} We consider a GHZ state with one additional qubit and the qubit in set $A$ is not affected by noise and later measured out. In this case the
	noise can be described by local noise (section \ref{sec:binlikemix_ghz}) and while such a noise-free virtual particle is clearly un-physical,
	it stands to reason that with the symmetry of the GHZ state an additional qubit should not affect the overall scaling.

\item \textit{Scenario C:} We only consider the decoding part and the threshold in this case is the value where there remains an interval of external noise
	for which it is advantageous to encode the information sent. This basically extends the approach of section \ref{sec:ghzvariant_extnoise} to a higher number of encoding qubits.
\end{itemize}

\begin{figure}
 \centering
 \includegraphics[width=\columnwidth]{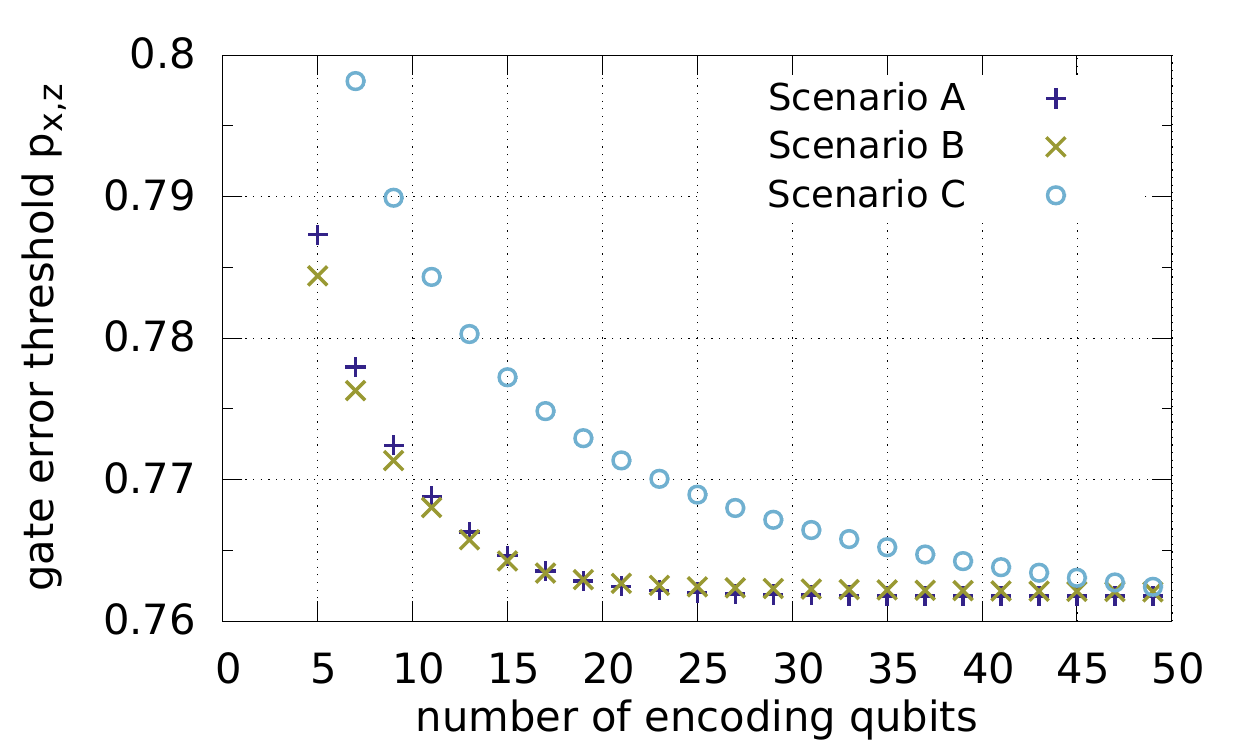}
 \caption{\label{fig:binary_thresh} Scaling of the gate noise threshold $p_{x,z}$ with the number of encoding qubits in the  approximate scenarios described in the main text.}
\end{figure}

In figure \ref{fig:binary_thresh} the scaling of the thresholds is depicted for these scenarios and we find that all of those tend towards a value of $p_{x,z} \approx 0.762$. That is, these results indicate that a quantum memory that allows one to protect quantum information against bit-flip or phase-flip errors can be obtained with such a measurement-based approach up to this error threshold.

\section{Summary and Outlook\label{sec:summary}}
 We have shown that using a local noise model for resource states in measurement based quantum information processing is justified. While not exact, such a description serves as a good approximation if the resource states are generated by entanglement purification. For certain resource states, such as GHZ states that are of importance for encoded computations using repetition codes, we have shown that an exact description by a local noise model is possible. This is done by further processing the resulting state locally and slightly adding noise thereby reducing the fidelity of the state by a small amount. It would be interesting to see if such a result can be generalized to other relevant resource states.

 Furthermore, we found that the resource states generated via EPP perform well when used for quantum error correction for a large range of parameters, and the locality of the noise does not appear to be a central property for this application. The details of the preparation procedure can influence the performance of the resource states significantly, so it is worthwhile to optimize it for specific applications.

 Finally, we have started to relate error models for measurement-based implementation to gate-based error models for specific tasks and a simplified error model. It would be interesting to generalize this analysis to fault-tolerant quantum computation with full noise, however it seems that novel techniques and methods are required in this case.

 \section*{Acknowledgements \label{sec:ack}}
 This work was supported by the Austrian Science Fund (FWF): P24273-N16, P28000-N27.

 \bibliography{literature}
\appendix

\section{\label{sec:On} Linear number of coefficients for binary-like mixture GHZ states}
The type of density matrices one has to consider for the specific variant of binary-like mixtures of GHZ states as described in the main text are given by
  \begin{equation}
    \rho=\sum_{\bm{k} \in \{0,1\}^{N_b}} c_{\bm{k}} \Sketbra{0,\bm{k}}{0,\bm{k}}{G} \text{,}
  \end{equation}
where $N_b = N - 1$ is the number of qubits in set $B$.

We now use symmetries to further reduce the number of parameters. The initial state one starts out with is expected to be symmetric with respect to permutation of particles in set $B$
because the underlying graph is symmetric in this regard. That means that $c_{\bm{k}} = c_k$ for all binary strings $\bm{k}$ with
the same number of ones $k=|\bm{k}|$, so only $N$ different coefficients have to be tracked. All operations used will preserve
this symmetry, but to actually benefit from this computationally it is necessary to reformulate everything with respect to these
coefficients.

Applying a local $\sigma_z$-noise channel on every qubit in set $B$ can be written as
\begin{widetext}
\begin{equation}
  \mathcal{E} \sum_{\bm{k} \in \{0,1\}^{N_b}} c_k \Sketbra{0,\bm{k}}{0,\bm{k}}{G}
  = \sum_{\bm{k} \in \{0,1\}^{N_b}} \sum_{\bm{j} \in \{0,1\}^{N_b}} c_k p^{N_b-j} (1-p)^j \Sketbra{0,\bm{k}\oplus\bm{j}}{0,\bm{k}\oplus\bm{j}}{G} =
									 \sum_{\bm{l} \in \{0,1\}^{N_b}} \widetilde{c}_l \Sketbra{0,\bm{l}}{0,\bm{l}}{G}
\end{equation}
\end{widetext}
with $\mathcal{E} = \prod_{i \in B} \mathcal{D}^i_z (p)$. After some calculation we find for the updated coefficients $\widetilde{c}_l$
  \begin{equation}
  \widetilde{c}_l = \sum_{k=0}^{N_b} c_k p^{N_b-k} (1-p)^k \sum_{i=0}^l \binom{l}{i} \binom{N_b-l}{k-i} p^{-l+2i} (1-p)^{l-2i} \text{.}
  \end{equation}
The effect of the $\sigma_x$ noise on set $A$ is easier to formulate:
\begin{equation}
 \widetilde{c}_l= p c_l + (1-p) c_{N_b-l} \text{.}
\end{equation}
For the entanglement purification only one of the two subprotocol is needed for the binary
like mixture and its effect is given by squaring the coefficients and renormalizing.
Putting this into this framework leads to
\begin{equation}
 \widetilde{c}_l=\frac{c_l^2}{\sum_k \binom{N_b}{k} c_k^2 } \text{.}
\end{equation}
In this way, one can treat such symmetric binary-like mixtures up to much larger particle numbers, and investigate the performance of multiparty EPP, locality of noise and the usage of resulting states for measurement-based communication.

\section{\label{sec:lcrule} Local complementation and graph state basis transformation}
 Two graphs $G$ and $G'$ are called local unitary (LU) equivalent if there exists a local unitary $U$
 such that $U\Ket{G} = \Ket{G'}$. One special way how two graphs can be shown to be LU-equivalent is the
 \textit{local complementation rule} \cite{graphstates}. The local complementation of a graph with respect to vertex $a$ is
 obtained by inverting the subgraph of all vertices in the neighborhood of $a$: $N_a = \{ b \mid {a,b} \in E \}$
  \begin{equation}
   \tau_a: G=(V,E) \mapsto \tau_a(G) = (V,E') \text{,}
  \end{equation}
  where
  \begin{equation}
   E'=E\triangle \{ \{b,c\} \in [V]^2 \mid b \in N_a , c \in N_a, b \neq c \}
  \end{equation}
  and $\triangle$ denotes the symmetric difference $E\triangle F= \left( E \cup F \right) - \left( E \cap F \right)$.

  The graph state associated with the graph $\tau_a(G)$ can be obtained from $\Ket{G}$ by applying
  local Clifford operations
  \begin{equation}
  \begin{aligned}
   \Ket{G'} &= \Ket{\tau_a (G)} = U_a^\tau (G) \Ket{G} \; \textrm{with}  \\
   U_a^\tau (G) &= \sqrt{-i \sigma_x^a} \prod_{j \in N_a} \sqrt{i \sigma_z^j} \propto \sqrt{K_a} \text{.}
  \end{aligned}
  \end{equation}
  From this rule we can calculate how the coefficients given in the graph state basis with respect to $G$ transform
  to the graph state basis with respect to $G'$. That is to say we look for $\Ket{\bm{\mu}'}_{G'}$ that fulfil
  \begin{equation}
   \Ket{\bm{\mu}'}_{G'} = U_a^\tau (G) \Ket{\bm{\mu}}_G \; \text{and} \; K_b' \Ket{\bm{\mu}'}_{G'} = (-1)^{\mu_b'} \Ket{\bm{\mu}'}_{G'}
  \end{equation}
  with $K_b' = \sigma_x^b \sigma_z^{N_b'}$ being the correlation operator around $b$ corresponding to the new graph $G'$ and with $N_b'$ as
  the neighborhood of $b$ with respect to $G'$.

  To calculate $\mu_b'$ we apply $K_b'$ on $\Ket{\bm{\mu}'}_{G'}$ and pay attention to the eigenvalue
  \begin{equation}
   K_b' \Ket{\bm{\mu}'}_{G'} = K_b' U_a^\tau (G) \ket{\bm{\mu}}_G \text{.}
  \end{equation}
  For $b \notin N_a$ the two operators $K_b'$ and $U_a^\tau (G)$ commute and we simply obtain: $K_b' \Ket{\bm{\mu}'}_{G'} = (-1)^{\mu_b} \Ket{\bm{\mu}'}_{G'}$ .
  The more interesting case $b \in N_a$ can be solved by using the relations
  \begin{equation}
   \begin{aligned}
    \sigma_x \sqrt{i \sigma_z} &= i \sqrt{i \sigma_z}  \sigma_x \sigma_z \\
    \sigma_z \sqrt{-i \sigma_x} &= (-i) \sqrt{-i \sigma_x} \sigma_z \sigma_x.
   \end{aligned}
   \label{eqn:sqrtperm}
  \end{equation}
  We obtain
  \begin{widetext}
   \begin{equation}
      \begin{aligned}
      & K_b' \Ket{\bm{\mu}'}_{G'} = K_b' U_a^\tau (G) \ket{\bm{\mu}}_G = \sigma_x^b \sigma_z^{N_b'} \sqrt{-i \sigma_x}^a \sqrt{i \sigma_z}^{N_a} \ket{\bm{\mu}}_G=
      \text{[using \eqref{eqn:sqrtperm}]} = \\ = &U_a^\tau (G) \sigma_x^b \sigma_z^b \sigma_z^{N_b'} \sigma_x^a \ket{\bm{\mu}}_G =
	  U_a^\tau (G) \sigma_x^b \sigma_z^b \sigma_z^{N_b'} \underbrace{\sigma_z^{N_a} \sigma_z^{N_a}}_{=\mathbbm{1}}\sigma_x^a \ket{\bm{\mu}}_G =
      U_a^\tau (G) \sigma_x^b \sigma_z^b \sigma_z^{N_b'} \sigma_z^{N_a} \underbrace{\sigma_z^{N_a} \sigma_x^a}_{=K_a} \ket{\bm{\mu}}_G = \\
	  = &(-1)^{\mu_a} U_a^\tau (G) \sigma_x^b \sigma_z^{N_b' \oplus (N_a \backslash {b})} \ket{\bm{\mu}}_G =
	  (-1)^{\mu_a} U_a^\tau (G) \underbrace{\sigma_x^b \sigma_z^{N_b}}_{=K_b} \ket{\bm{\mu}}_G = (-1)^{\mu_a \oplus \mu_b} \Ket{\bm{\mu}'}_{G'}.	
      \end{aligned}
   \end{equation}
  \end{widetext}

  To summarize, it holds that
  \begin{equation}
   \mu_b' = \begin{cases}
             \mu_a \oplus \mu_b  & \; \text{for } b \in N_a \\
             \mu_b & \; \text{otherwise}
            \end{cases}
  \end{equation}
  which means that some of the coefficients in the density matrix are simply exchanged.

\section{\label{sec:purdetails} Purification procedure for the cluster-ring code}
As explained in the main text the details for generating the resource states influence their performance dramatically.
For the cluster-ring code there are also more variables that have to be adjusted than for the GHZ states.

\begin{figure}[ht]
 \centering
 \subfloat{\includegraphics[width=0.25\columnwidth]{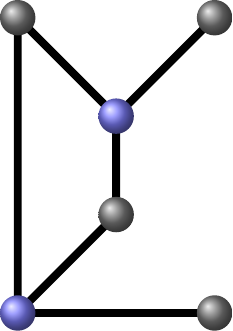}}
 \hfill
 \subfloat{\includegraphics[width=0.25\columnwidth]{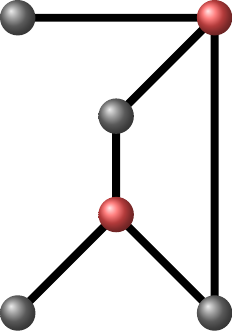}}
 \hfill
 \subfloat{\includegraphics[width=0.25\columnwidth]{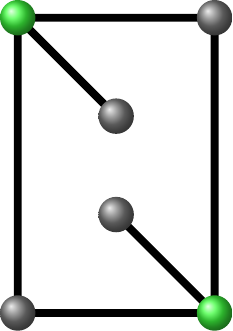}}
 \caption{\label{fig:aux_graphs} Auxiliary two-colorable graph states used for purifying the 6-qubit graph state in figure \ref{fig:lc_clr}. }
\end{figure}

It is necessary to use the EPP that works for all graph states \cite{eppallgraphs} since the resource state is not
two-colorable and needs four colors instead. The protocol is very similar to the two-colorable protocol \cite{adbepp}
described in the main part, however, a separate subprotocol $P_i$ is used for each color. Furthermore, it does not use two
copies of the same state at each step but needs auxiliary two-colorable graph states $g_i$ for each of the subprotocols.
We consider these auxiliary states to have been purified by the EPP for two-colorable graph states using CNOT games with
the same error parameter. It should be noted that alternatively these auxiliary states can also be prepared from the main
graph state \cite{eppallgraphs}.
The protocol works better for fewer colors, therefore we use a three-colorable state that is LU-equivalent to the resource
state for the cluster-ring code (see figures \ref{fig:lc_clr} and \ref{fig:aux_graphs} ). This state is already mentioned in \cite{clusterringcode}
but we do not use it for error correction directly as the cluster-ring is more intuitive. It is important to consider
that some coefficients are exchanged when switching from one graph state basis to one based on another graph as described in Appendix \ref{sec:lcrule}.

To optimize the preparation procedure not only do we have to optimize where to stop in the cycle of $P_1$-$P_2$-$P_3$-\dots for the main graph state,
but also whether to end the EPPs for the auxiliary (two-colorable) states with P1 or P2. As for the GHZ states this procedure could likely be optimized further.

\section{\label{sec:lcs} Linear Cluster state}
Here we investigate the noise structure of other graph states at their fixed point of the EPP.
 Linear cluster states have less symmetry that can be used to lower the amount of parameters one has
 to consider for the minimization process than for GHZ states. We first and foremost investigate the linear cluster
 state with $N=5$ qubits because it has a known application in a $4\rightarrow1$ measurement
 based EPP \cite{mbqc_eppthresh, mbqrepeaters} for bipartite entangled states.

\begin{figure}[ht]
  \includegraphics[width=\columnwidth]{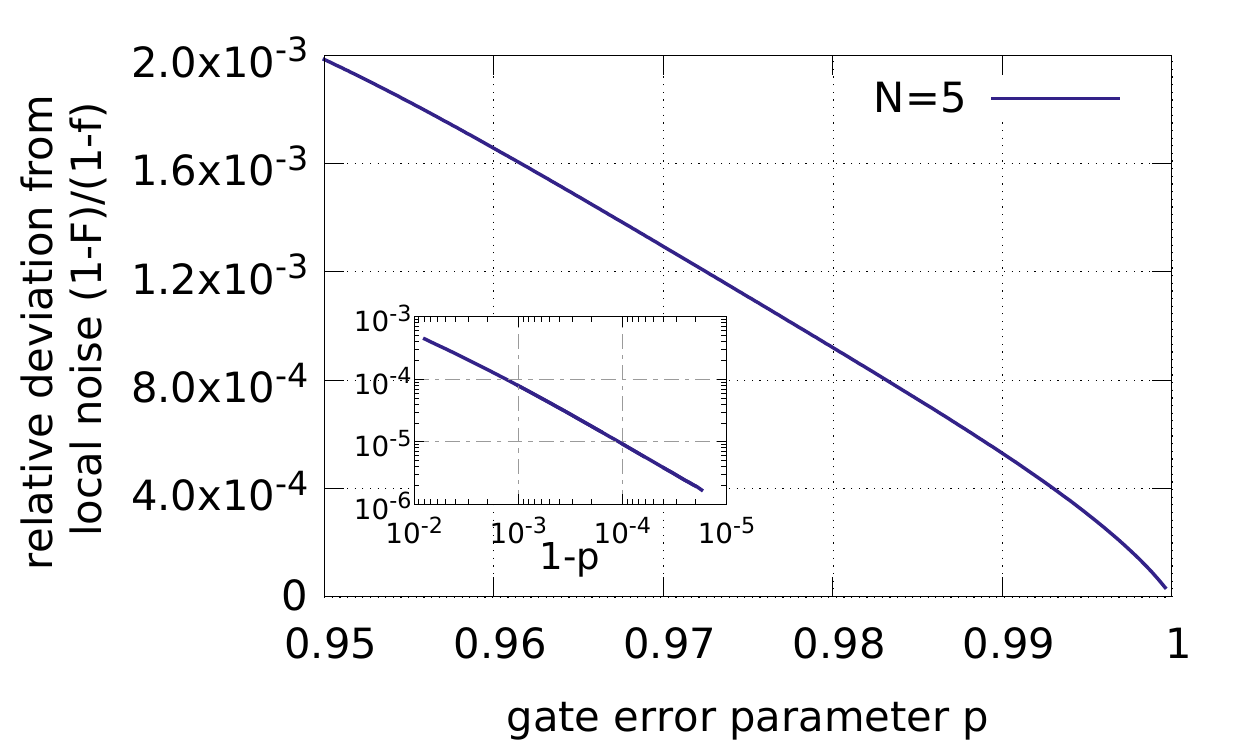}
  \caption{\label{fig:lc5_rel} Relative deviation from a local noise model at the fixed point of the multipartite EPP for the linear cluster state
	   with $N=5$ qubits. The gate error parameter $p$ refers to the CNOT gates used in the EPP. }
 \end{figure}

 For local white noise as gate error model we find a similar behavior for the deviation from a local noise model as for the GHZ states. While
 the deviation is bigger than for GHZ states of similar size, the local noise model still serves as a good
 approximation, especially for $p$ close to $1$ (figure \ref{fig:lc5_rel}).

\section{\label{sec:localizing_details} Localizing noise for GHZ states}
In \cite{Du00} it is shown that every state can be brought to the standard form of equation \eqref{eqn:ghzstandardform} in the main text
by the random application of local unitaries. One particular way to achieve this with local Clifford operations is to apply
$U_a = \sqrt{-i\sigma_x^{a} } \sqrt{i \sigma_z^1}$ with probability $1/2$ each for all $a \in \{2, \dots, N\}$.
This is the case because the application of $U_a$ exchanges graph state basis states as follows: $U_a \Ket{\bm{\mu}}_G = \Ket{\bm{\mu'}}_G$
with $\mu'_1 = \mu_1 \oplus \mu_a$ where $\oplus$ denotes the addition mod 2. This results in making $\lambda^{+}_{\bm{k}}$ and $\lambda^{-}_{\bm{k}}$
equal for $\bm{k} \neq \bm{0}$ and leaving $\lambda^\pm_{\bm{0}}$ unchanged.

In the main part it is already stated that the central trick in making the noise follow a local noise pattern is to
probabilistically add separable states to the ensemble
\begin{equation}
 \begin{aligned}
  \rho_{\bm{k}} = 1/2 \bigl( \Ketbra{\Psi^+_{\bm{k}} }{\Psi^+_{\bm{k}} } &+ \Ketbra{\Psi^-_{\bm{k}} }{\Psi^-_{\bm{k}} } \bigr) = \\
  1/2 U_{\bm{k}}[ \Ketbra{0+\dots+}{0+\dots+} &+ \Ketbra{1-\dots-}{1-\dots-} ]U_{\bm{k}}^\dagger
 \end{aligned}
\end{equation}
with $U_{\bm{k}} = \prod_{i=2}^N \left(\sigma_z^i \right)^{k_i}$ and $|\pm\rangle=(|0\rangle \pm |1\rangle)/\sqrt{2}$.
These states can easily be prepared and directly change the coefficients $\lambda_{\bm{k}}$ of the updated state
\begin{equation}
 \rho' = Q \rho + \sum_{\bm{k}} q_{\bm{k}} \rho_{\bm{k}}
\end{equation}
 with appropriate weights $q_{\bm{k}} > 0$, which have to satisfy: $Q = 1 -  \sum_{\bm{k}} q_{\bm{k}}$.

Next we identify the type of local noise that results in states in the standard form as noise channels with equal coefficients
for $\sigma_x$ and $\sigma_y$ noise on set A and equal coefficients for $\sigma_y$ and $\sigma_z$ noise on set B. Symmetries in the
GHZ state and the resulting state from entanglement purification make assigning equal weights to the noise on sets A and B optimal.

Furthermore, the ratio of $\lambda^+_{\bm{0}}$ and $\lambda^-_{\bm{0}}$ can be freely chosen with either $\sigma_z$ noise on set A
or $\sigma_x$ noise on set B so this ratio does not affect whether the noise can be described by a local noise model and it is enough
to consider $\widetilde{\lambda}_{\bm{0}} =  \lambda^+_{\bm{0}} + \lambda^-_{\bm{0}}$ and $\widetilde{\lambda}_{\bm{k}} = 2 \lambda_{\bm{k}}$
for $\bm{k} \neq \bm{0}$.

With the noise channels $\mathcal{D}_A^i(p) \rho = p \rho + \frac{1-p}{2} (\sigma^i_x \rho \sigma^i_x + \sigma^i_y \rho \sigma^i_y)$ for set A
and $\mathcal{D}_B^i(p) \rho = p \rho + \frac{1-p}{2} (\sigma^i_y \rho \sigma^i_y + \sigma^i_z \rho \sigma^i_z)$ for set B we get

\begin{equation}
 \widetilde{\lambda}'_{\bm{k}} = p^k (1-p)^{(N-1)-k} + p^{(N-1)-k} (1-p)^k \text{,}
\end{equation}
where $k = |\bm{k}|$ is the Hamming weight of $\bm{k}$.

Because we want the fidelity $\lambda^+_{\bm{0}}$ to stay as high as possible while matching the coefficients given
by the local noise model with parameter $p$ and it is only possible to modify $\lambda^+_{\bm{0}}$ and $\lambda^-_{\bm{0}}$
together, the challenge lies in picking $Q$ as large as possible and use the according $q_{\bm{k}}$ . This is achieved by

\begin{equation}
 Q = \max_{\bm{k}} \frac{\widetilde{\lambda}'_{\bm{k}}}{\widetilde{\lambda}_{\bm{k}} } \quad ; \quad q_{\bm{k}}=\widetilde{\lambda}'_{\bm{k}} - Q \widetilde{\lambda}_{\bm{k}} \text{,}
\end{equation}
where $\widetilde{\lambda}'_{\bm{k}}$ are the coefficients of $\rho'$. The outcome of this process is that we obtain a function $Q(p)$.

So far everything can be done analytically, but in the final step we numerically maximize $Q(p)$
which results in the fidelity of $\rho'$ being as high as possible.

\section{Circuits for gate-based error correction \label{sec:qeccircuit}}
 In the main text we compare different implementations of quantum error correction codes. For the gate-based implementation of the well-known bit-flip code
 we use the standard circuit found e.g. in \cite{niechu}.
 \begin{figure}
  \includegraphics[width=\columnwidth]{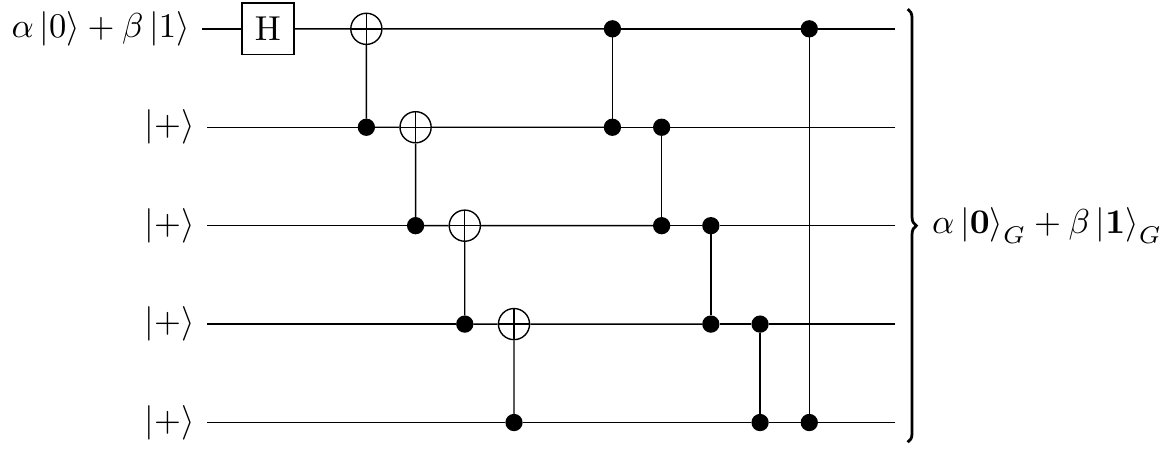}
  \caption{ Circuit for encoding a state using the cluster-ring code. The operations used are a Hadamard gate, four CNOT gates followed by a pattern of five CZ gates. \label{fig:clrcircuit} }
 \end{figure}
 In figure \ref{fig:clrcircuit} the circuit we use for encoding the cluster-ring code is shown. The circuit is not unique and most likely not optimal.
 For decoding the circuit is performed in reverse order and the last four qubits are measured in the $\sigma_x$ basis to obtain the error syndrom and make appropriate corrections.

\section{\label{sec:dec_patterns} Measurement based decoding patterns}
 In order to analyze the performance of resource states for error correction codes in the measurement based approach it is necessary to formulate
 the full protocol with all correction operations because with an error model that cannot be described by local noise it is not possible to simply
 look at a noise channel acting on the input followed by the perfect protocol. We will discuss this process in detail for the bit-flip variant of the
 repetition code, but it can easily be extended to more involved codes like the cluster-ring code or slightly reformulated to fit the phase-flip variant
 (for which the resource state is given directly by the graph state in figure \ref{fig:ghz4} in the main text).

 When decoding an encoded quantum state with a resource state the measurement outcomes of the Bell measurements contain all the information to identify
 a potential error that has occurred. One way to understand what correction operations the different outcome patterns imply is to simply consider the
 way the resource state is calculated using the Jamiolkowski isomorphism \cite{jamiolkowski} for the circuit that implements the decoding. The
 qubits that are measured in the circuit model can be interpreted as virtual qubits that are disconnected from the resource state by the measurement.
 Then, one can calculate what effect the byproduct operators from the different Bell measurement outcomes have on the measurement outcomes on these virtual qubits
 and simply use the correction operators one would use for the circuit implementation.

 Alternatively, it is possible to understand the patterns without referring to a circuit. A projection on the Bell state $\Ket{\Phi_i} = \mathbbm{1} \otimes \sigma_i \Ket{\Phi^+}$
 can be understood as applying $\sigma_i$ on the input followed by projecting on $\Ket{\Phi^+}$ for which we know it implements the decoding without any additional byproduct operators,
 so the measurement outcomes can be interpreted as modifying the input only. In case no error happened, only combinations of $\sigma_i$ that leave the subspace spanned by
 $\Ket{0_L}$ and $\Ket{1_L}$ invariant will have a non-zero probability of occurring. In table \ref{tab:patterns} all combinations that can occur if no error happened are listed
 together with the resulting output state and the necessary correction operations if the input state is $\alpha \Ket{0_L} + \beta \Ket{1_L}$. Similarly, the outcomes which
 indicate that an error occurred are the ones that map it to the subspace corresponding to that error.

 \begin{table}[ht]
  \begin{tabular}{ccc|cc}
   $\sigma_{i_1}$ & $\sigma_{i_2}$ & $\sigma_{i_3}$ & $\Ket{\Psi_\mathrm{out}}$ & correction \\
   \hline
   $\mathbbm{1}$ & $\mathbbm{1}$ & $\mathbbm{1}$ & $\alpha \Ket{0} + \beta \Ket{1}$ & $\mathbbm{1}$ \\
   $\sigma_z$ & $\sigma_z$ & $\mathbbm{1}$ & $\alpha \Ket{0} + \beta \Ket{1}$ & $\mathbbm{1}$ \\
   $\sigma_z$ & $\mathbbm{1}$ & $\sigma_z$ & $\alpha \Ket{0} + \beta \Ket{1}$ & $\mathbbm{1}$ \\
   $\mathbbm{1}$ & $\sigma_z$ & $\sigma_z$ & $\alpha \Ket{0} + \beta \Ket{1}$ & $\mathbbm{1}$ \\
   $\sigma_z$ & $\mathbbm{1}$ & $\mathbbm{1}$ & $\alpha \Ket{0} - \beta \Ket{1}$ & $\sigma_z$ \\
   $\mathbbm{1}$ & $\sigma_z$ & $\mathbbm{1}$ & $\alpha \Ket{0} - \beta \Ket{1}$ & $\sigma_z$ \\
   $\mathbbm{1}$ & $\mathbbm{1}$ & $\sigma_z$ & $\alpha \Ket{0} - \beta \Ket{1}$ & $\sigma_z$ \\
   $\sigma_z$ & $\sigma_z$ & $\sigma_z$ & $\alpha \Ket{0} - \beta \Ket{1}$ & $\sigma_z$ \\
   $\sigma_x$ & $\sigma_x$ & $\sigma_x$ & $\alpha \Ket{1} + \beta \Ket{0}$ & $\sigma_x$ \\
   $\sigma_y$ & $\sigma_y$ & $\sigma_x$ & $\alpha \Ket{1} + \beta \Ket{0}$ & $\sigma_x$ \\
   $\sigma_y$ & $\sigma_x$ & $\sigma_y$ & $\alpha \Ket{1} + \beta \Ket{0}$ & $\sigma_x$ \\
   $\sigma_x$ & $\sigma_y$ & $\sigma_y$ & $\alpha \Ket{1} + \beta \Ket{0}$ & $\sigma_x$ \\
   $\sigma_y$ & $\sigma_x$ & $\sigma_x$ & $\alpha \Ket{1} - \beta \Ket{0}$ & $\sigma_y$ \\
   $\sigma_x$ & $\sigma_y$ & $\sigma_x$ & $\alpha \Ket{1} - \beta \Ket{0}$ & $\sigma_y$ \\
   $\sigma_x$ & $\sigma_x$ & $\sigma_y$ & $\alpha \Ket{1} - \beta \Ket{0}$ & $\sigma_y$ \\
   $\sigma_y$ & $\sigma_y$ & $\sigma_y$ & $\alpha \Ket{1} - \beta \Ket{0}$ & $\sigma_y$
  \end{tabular}
  \caption{\label{tab:patterns} Bell measurement outcome patterns and correction operations for the measurement based repetition code decoding if no error occurred for the
	  input state $\Ket{\Psi_\mathrm{in} } = \alpha \Ket{0_L} + \beta \Ket{1_L}$ .}
 \end{table}

 The same intuition can also be applied for more complex codes - the patterns corresponding to no error occuring are those combinations of Pauli operators that
 leave the subspace spanned by $\Ket{0_L}$ and $\Ket{1_L}$ invariant. These select tensor products of Pauli operators form a subgroup of the stabilizer of $\Ket{0_L}$.
 The patterns of certain error are then
 obtained by simply multiplying the error operator with the elements of this group.

 Furthermore, it should be stressed that the error codes are not only able to correct errors on the input but also noise on the resource states without any modifications to
 the decoding procedure. Because $\mathbbm{1} \otimes \sigma_i \Ket{\Phi^+} \propto \sigma_i \otimes \mathbbm{1} \Ket{\Phi^+}$ the patterns for an error are the
 same as if they occurred on the input.

 For practical purposes (at least for small resource states) these patterns can be found numerically in a straightforward way. Simply apply an error on the input, for which we know that the code can correct it, and perform the read-in to a perfect resource state. Then, one can see which of the outcome patterns occur for this type of error and immediately obtain the correction operation by checking what the output state is. Doing the same for all the errors the code can detect, one obtains a table with all of the patterns and the corresponding correction operations.

\end{document}